\documentclass[letter,10pt]{article}
\usepackage{mathptmx,graphicx,eprint}

\setreportheading{Masahiko Yoshioka}{Preprint}

\begin{document}

{\title{{Linear stability analysis of retrieval state in associative memory neural networks of spiking neurons}}}
{%
\author{{{Masahiko Yoshioka\thanks{Electronic address: myosioka@brain.riken.go.jp}}}\\
{\small\it{{Brain Science Institute, RIKEN}}}\\
{\small\it{{Hirosawa 2-1, Wako-shi, Saitama, 351-0198, Japan}}}}%
\date{{{August 5, 2002}}\\
{\small{{(Revised on September 19, 2002)}}}}%
\maketitle}

\begin{abstract}
We study associative memory neural networks of the Hodgkin-Huxley type of spiking neurons in which multiple periodic spatio-temporal patterns of spike timing are memorized as limit-cycle-type attractors.
In encoding the spatio-temporal patterns, we assume the spike-timing-dependent synaptic plasticity with the asymmetric time window.
Analysis for periodic solution of retrieval state reveals that if the area of the negative part of the time window is equivalent to the positive part, then crosstalk among encoded patterns vanishes.
Phase transition due to the loss of the stability of periodic solution is observed when we assume fast $\alpha$-function for direct interaction among neurons.
In order to evaluate the critical point of this phase transition, we employ Floquet theory in which the stability problem of the infinite number of spiking neurons interacting with $\alpha$-function is reduced into the eigenvalue problem  with the finite size of matrix.
Numerical integration of the single-body dynamics yields the explicit value of the matrix, which enables us to determine the critical point of the phase transition with a high degree of precision.
\end{abstract}

\section{Introduction}

Synchronized firing of neurons are ubiquitous phenomena in real nervous system, and capability of synchronicity of neurons for information processing has been the subject of many research papers\cite{gray,eckhorn,malsburg,kuramoto,vreeswijk,gerstner,kopell,yoshioka3,yoshioka4,myosioka5,hasegawa}.
It has been revealed that repeating firing patterns of pyramidal neurons appear in sharp waves of rat hippocampus\cite{nadasdy}.
This result of experiment suggests the possible role of spatio-temporal patterns of spike timing in encoding information in real nervous system.
Associative memory neural networks that memorize spatio-temporal patterns of spike timing is essential for understanding this information processing of spike timing.

Much of fundamental concepts of associative memory neural networks have been developed by replica calculation of Ising spin neural networks with the energy function\cite{hopfield,amit5,amit2}.
In these neural networks, the standard type of Hebb rule is assumed to define symmetric synaptic connections, which bring about fixed-point-type attractors.
These fixed-point-type attractors are, however, useless for encoding spatio-temporal patterns.
 Asymmetric synaptic connections play a significant role in encoding spatio-temporal patterns, and then the question arises about the learning rule that defines asymmetric synaptic connections so that the network functions as associative memory for sptio-temporal patterns.
When we assume synchronous update rule for the dynamics of spin neural networks, a simple extension of the Hebb rule readily realizes associative memory for spaio-temporal patterns\cite{amari3}.
Nevertheless, the problem becomes rather difficult when we assume asynchrnous update rule for spin neural networks.
Complicated learning rules are required to control the continuous transition of network state in sequential retrieval of spatial patterns\cite{nishimori,sompolinsky2}.

In spin neural networks\cite{amari2,gardner,coolen}, as well as analog neural networks\cite{kuhn,kuhn2,shiino,shiino2,shiino3,okada_n2,yoshioka,yoshioka_n2}, state variables of neurons are assumed to represent their firing rate.
In neural networks of phase oscillators,  phase variables are used to represent synchrnoized firing of neurons.
The synaptic connections of Hermitian permits networks of oscillators to memorize spatio-temporal patterns of phase differences.
Since some theoretical techniques are available for the analysis of phase oscillators, the properties of networks of oscillators have been investigated extensively\cite{sakaguchi,arenas,hansel2,cook,aoyagi,yamana}.
Even in the presence of white noise as well as heterogeneity of oscillators we can derive the storage capacity of networks of oscillators analytically\cite{yoshioka4}.

Nevertheless, networks of oscillators may possibly offer a distorted interpretation of synchronized firing in the real nervous system unless interactions among neurons are sufficiently weak.
To provide a real understanding of the information processing of spike timing, we must adopt more biologically plausible models of neural networks.
For this purpose, neural networks of spiking neurons are considered to be suitable models for investigation, though it remains an unsolved problem to find adequate learning rule for spatio-temporal patterns of spike timing.
Since asymmetric synaptic connections bring about sequential firings of spiking neurons\cite{wallenstein,tsodyks}, one may consider that asymmetric synaptic connetions are essential for associative memory neural networks of spiking neurons.
In fact, incorporating asymmetric synaptic connections, Gerstner {et al}. have successed in encoding a few number of spatio-temporal patterns in networks of spiking neurons with discrete time dynamics\cite{gerstner}.

\begin{figure}
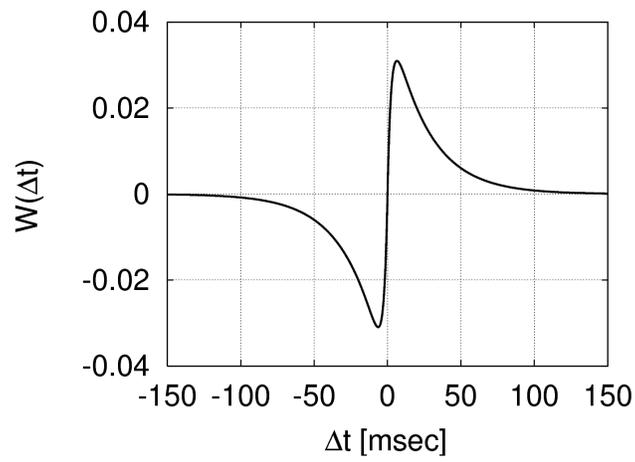

{\begin{center}
{\includegraphics[scale={0.7}]{{{w}}}}
\end{center}}
\caption{The shape of the time window~$W{\left ( {\Delta t} \right )}$ with $\tau_{{{\rm W}},1}=25{\mbox{ [msec]}}$ and $\tau_{{{\rm W}},2}=2.5{\mbox{ [msec]}}$.
The modification of synaptic weight is written as $\Delta J\propto W{\left ( {\Delta t} \right )}$, where $\Delta t=t_{\rm post}-t_{\rm pre}$ denotes the time difference between the postsynaptic and presynaptic firings.}{\label{figw}}
\end{figure}

The spike-timing-dependent synaptic plasticity found in electrophysiological experiments excites a good deal of interest in this connection.
It has been revealed that the modification of excitatory synaptic weight depends on the precise timings of presynaptic and postsynaptic firings\cite{markram,guo,zhang}.
Synaptic weight is found to increase if firing of a presynaptic neuron occurs in advance of firing of a postsynaptic neuron, and to decrease otherwise.
The spike-timing-dependent synaptic plasticity is described by the time window with the negative part as well as the positive part ({{{Fig.}~\ref{figw}}}), and this asymmetric shape of the time window has been attracting a growing interest of reseachers\cite{abbott,gerstner3,mehta,mehta2,song,kempter,rubin,matsumoto}.
Since the asymmetric time window brings about asymmetric synaptic connections, the spike-timing-dependent synaptic plasticity is thought to be advantageous to encode spatio-temporal patterns.
In the previous study of ours, we have studied associative memory neural networks of spiking neurons in which the asymmetric time window of the spike-timing-dependent synaptic plasticity is used to encode multiple periodic spatio-temporal patterns of spike timing\cite{myosioka5}.
We have assumed networks of the Hodgkin-Huxley neurons interacting through direct synaptic interaction, as well as indirect synaptic interaction intermediated by firings of interneurons.
In the process of memory retrieval, the indirect interactions bring about the oscillatory inhibitory electric currents, which regulate spike timings of neurons as in the case of gamma and ripple oscillations\cite{bragin,ylinen,buzsaki2}.
In order to elucidate the stationary properties of these retrieval state we have derived the periodic solution for retrieval state analytically, and then we have shown that if the area of the negative part of the time window is equivalent to the positive part, crosstalk among encoded patterns vanishes.
This theoretical result implies the outstanding nature of the spike-timing-dependent synaptic plasticity for encoding multiple spatio-temporal patterns.

In our previous study, however, we did not carry out a stability analysis for the retrieval state, and hence it remained unclear whether the derived retrieval state are stable or not.
We investigate the same models of neural networks also in the present study, but we assume that the $\alpha$-function of the direct interaction decays much faster than the previous one.
Then, we find the phase transition due to the loss of the stability of retrieval state.
In order to evaluate the critical point of this phase transition we consider employing Floquet thoery.
Neverthless, the degree of freedom of the present system is infinite, and the naive application of Floquet theory yields the eigenvalue problem with the infinite size of matrix.
Furtheremore, $\alpha$-function we assume here exhibits the infinite long-time influence, and its treatment may also require the infinite size of matrix\cite{gerstner4}.
Without calculating the explicit form of the matrix, Bressloff {et al}. have investigated the stability of some periodic solutions in networks of integrate-and-fire neurons\cite{bressloff3}, though its application to other neuron models, such as the FitzHugh-Nagumo neurons and the Hodgkin-Huxley neurons, seems to be limited.

In the present study, we employ two theoretical techniques to reduce the size of the matrix for Floquet theory.
We first take the limit of the infinite large number of neurons, which reduces the stability problem of $N$ neurons into the problem of $Q$ sublattices.
Then, we define some additional variables for each sublattice and evalute the infinite long-time influence of $\alpha$-function with the finite size of the matrix.
The explicit value of the matrix is calculated from the numerical integration of the single-body dynamics.
Therefore, we can explicitly obtain the eigenvalues of the matrix, which enable us to determine the critical point of the phase transition with a high degree of precision, even when we assume networks of Hodgkin-Huxley neurons.

The present paper is organized as follows.
In {{{section}~\ref{sect:associativememory}}}, we present the dynamics of neural networks of spiking neurons, and then introduce the spike-timing-dependent learning rule for associative memory.
In {{{section}~\ref{sect:stationarystate}}}, we derive the retrieval state analytically in the limit of the infinite number of neurons.
After that, the stability of this retrieval state is analyzed by Floquet theory in {{{section}~\ref{sect:stabilityanalysis}}}.
In {{{section}~\ref{sect:retrievalprocess}}}, we illustrate the typical behavior of network in the process of memory retrieval.
The result of numerical simulations are presented, and compared with the theoretical results.
In {{{section}~\ref{sect:lossofstability}}}, we discuss the phase transition due to the loss of the stability based on the stability analysis in {{{section}~\ref{sect:stabilityanalysis}}}.
In {{{section}~\ref{sect:separated}}}, we investigate the case of slow $\alpha$-function, with which we find two separated retrieval phases.
In one of these retrieval phases, neurons obtain the large size of the oscillatory inhibitory synaptic electric currents, which well regulate the spike timing of neurons.
Finally, in {{{section}~\ref{sect:discussion}}}, we give a brief summary and discuss the biological implication of the present study.

\section{Associative memory neural networks of spiking neurons}{\label{sect:associativememory}}

\subsection{Network dynamics}

In real nervous system, many regions such as the neocortex and hippocampus are found to comprise a large number of pyramidal neurons as well as interneurons.
Our interest in the present study lies in spike timing of pyramidal neurons, and we denote the dynamics of $N$ pyramidal neurons by a set of nonlinear differentail equations of the form
\begin{eqnarray}
\dot{v}_i&=&f{\left ( {v_i,w_{i1},\ldots,w_{in}} \right )}+I_i,{\label{vdyn}}\\
\dot{w}_{il}&=&g_l{\left ( {v_i,w_{i1},\ldots,w_{in}} \right )},\qquad l=1,\ldots,n,\ i=1,\ldots,N {\label{wdyn}}
\end{eqnarray}
with
\begin{equation}
 I_i=I_{{{\rm PP}},i}+I_{{{\rm IP}}}+I_{{{\rm EXT}},i},{\label{isum}}
\end{equation}
where $v_i$ denotes membrane potential and auxiliary variables $w_{il}$ are used to describe gating of ion channels.
Synaptic electric currents $I_i$ denote interaction among neurons, and the definitions of three currents included in $I_i$ will be given later.
For the dynamics $f{\left ( {v_i,w_{i1},\ldots,w_{in}} \right )}$ and $g_l{\left ( {v_i,w_{i1},\ldots,w_{in}} \right )}$, many authors assume the integrate and fire equation, the FitzHugh-Nagumo equations\cite{fitzhugh,nagumo}, the Hodgkin-Huxley equations\cite{hodgkin}, and so on.
In the present study we choose the Hodgkin-Huxley equations, which are summarized in Appendix~\ref{hhequations}.

The synaptic electric current $I_{{{\rm PP}},i}$ in $I_i$ expresses the direct interaction among pyramidal neurons.
We define spike timing of neuron~$i$ as the time when the membrane potential $v_i$ exceeds the threshold value $\theta=0$.
Denoting $k$-th spike timing of neuron~$i$ by $t_i(k)$, we define the current $I_{{{\rm PP}},i}$ as
\begin{equation}
 I_{{{\rm PP}},i}=A_{{{\rm PP}}}\sum_{j=1}^N J_{ij}\sum_k S_{{{\rm PP}}}{\left [ {t-t_j(k)} \right ]},{\label{ipp}}
\end{equation}
where $J_{ij}$ represents the synaptic weight, and $\alpha$-function $S_{{{\rm PP}}}(t)$ is defined as
\begin{equation}
 S_{{{\rm PP}}}(t)=\left\{\begin{array}{ll} 
0 &{\mbox{for}\ \ } t<0,\\
\displaystyle\frac{1}{\tau_{{{\rm PP}},1}-\tau_{{{\rm PP}},2}}{\left ( {{e}^{-t/\tau_{{{\rm PP}},1}}-{e}^{-t/\tau_{{{\rm PP}},2}}} \right )} &{\mbox{for}\ \ } 0\le t.
\end{array}\right .{\label{spp}}
\end{equation}
The constant $A_{{{\rm PP}}}$ is used to control the intensity of the current $I_{{{\rm PP}},i}$.
In the following section, we will investigate the case of the fast $\alpha$-function~$S_{{{\rm PP}}}(t)$ with $\tau_{{{\rm PP}},1}=3{\mbox{ [msec]}}$ and $\tau_{{{\rm PP}},2}=0.3{\mbox{ [msec]}}$ as well as the slow $\alpha$-function~$S_{{{\rm PP}}}(t)$ with $\tau_{{{\rm PP}},1}=20{\mbox{ [msec]}}$ and $\tau_{{{\rm PP}},2}=2{\mbox{ [msec]}}$.

The synaptic electric current $I_{{{\rm IP}}}$ in $I_i$ expresses the indirect interaction among pyramidal neurons that is intermediated by firings of interneurons.
Since the threshold value for firing of interneurons is rather small, we assume that when one pyramidal neuron fires, interneurons surrounding the firing pyramidal neuron immediately fire.
Then, these firings of interneurons bring about inhibitory synaptic electric currents in all pyramidal neurons, because interneurons are connected to pyramidal neurons via inhibitory synapses.
This inhibitory synaptic electric current $I_{{{\rm IP}}}$, which is independent of index $i$, is written as
\begin{equation}
 I_{{{\rm IP}}}= A_{{{\rm IP}}}\sum_j K\sum_k S_{{{\rm IP}}}{\left [ {t-t_j(k)} \right ]},{\label{iip}}
\end{equation}
where $\alpha$-function $S_{{{\rm IP}}}(t)$ is defined as
\begin{equation}
 S_{{{\rm IP}}}(t)=\left\{\begin{array}{ll}
0 &{\mbox{for}\ \ } t<0,\\
\displaystyle\frac{-1}{\tau_{{{\rm IP}},1}-\tau_{{{\rm IP}},2}}{\left ( {{e}^{-t/\tau_{{{\rm IP}},1}}-{e}^{-t/\tau_{{{\rm IP}},2}}} \right )} &{\mbox{for}\ \ } 0\le t.
\end{array}\right .{\label{sip}}
\end{equation}
We set $K=1/N$ for proper scaling.
The constant $A_{{{\rm IP}}}$ is used to control the intensity of $I_{{{\rm IP}}}$, and the constants $\tau_{{{\rm IP}},1}$ and $\tau_{{{\rm IP}},2}$ are always taken to be $10 {\mbox{ [msec]}}$ and $1{\mbox{ [msec]}}$, respectively.
The function $S_{{{\rm IP}}}(t)$ takes negative value so as to represent the inhibitory nature of the connection.
Once note that we neglect the detailed dynamics of interneurons and simply assume that the inhibitory currents $I_{{{\rm IP}}}$ are induced in all pyramidal neurons immediately after one pyramidal neurons fires\cite{jensen}.

The current $I_{{{\rm EXT}},i}$ in $I_i$ is used to control initial firings of neurons.
For the initial condition of the network, we set all state of neuron ${\left ( {v_i,{\left\{ {w_{il}} \right \}}} \right )}$ to be at the stable fixed point of the dynamics~{{(\ref{vdyn})}} and {{(\ref{wdyn})}} with $I_i=0$. 
Since all neurons keep quiescent without any external stimuli, we use the pulsed form of the external electric current $I_{{{\rm EXT}},i}$ to invoke initial firings, which act as a trigger for information processing of the present model.
The detailed definition of $I_{{{\rm EXT}},i}$ will be given in {{{section}~\ref{sect:retrievalprocess}}}.
Note that the current $I_{{{\rm EXT}},i}$ is applied only in the beginning of the dynamics~{{(\ref{vdyn})}}-{{(\ref{wdyn})}}.
In the theoretical analysis below we always set $I_{{{\rm EXT}},i}=0$ because we focus on the stationary state in this analysis.

\subsection{Spike-timing-dependent learning rule}

We investigate associative memory neural network models that memorize multiple periodic spatio-temporal patterns of spike timing.
$P$ periodic spatio-temporal patterns to be memorized are generated randomly according to the equation
\begin{equation}
 \tilde{s}_i^\mu=s_i^\mu+kT,\qquad k={\ldots,-2,-1,0,1,2\ldots},\ \mu=1,\ldots,P,\ i=1,\ldots,N, {\label{pattern}}
\end{equation}
with
\begin{equation}
 s_i^\mu=\frac{T}{Q}q_i^\mu,
\end{equation}
where $Q$ is a natural number controlling the degree of discreteness of spatio-temporal patterns, and random integer $q_i^\mu$ is chosen from the interval $[0,Q)$ with equal probability.
$T$ denotes the period of the spatio-temporal patterns.
We set $T=250{\mbox{ [msec]}}$ and $Q=10$ in what follows.

Let us consider the problem of encoding the spatio-temporal patterns~$\tilde{s}_i^\mu$ so that the networks function as associative memory.
The recent results of the electrophysiological experiments have revealed that the modification of a synaptic weight depends on the precise timing of presynaptic and postsynaptic spikes\cite{markram,guo,zhang}.
Such modification of synaptic weight $\Delta J$ is approximately written in the form
\begin{eqnarray}
 \Delta J&\propto& W{\left ( {\Delta t} \right )}\nonumber\\
&=&\left\{\begin{array}{lc}
        \displaystyle \frac{-1}{\tau_{{{\rm W}},1}-\tau_{{{\rm W}},2}}{\left ( {{e}^{\Delta t/\tau_{{{\rm W}},1}}-{e}^{\Delta t/\tau_{{{\rm W}},2}}} \right )} &{\mbox{for}\ \ }  \Delta t<0,\\
        \displaystyle \frac{1}{\tau_{{{\rm W}},1}-\tau_{{{\rm W}},2}}{\left ( {{e}^{-\Delta t/\tau_{{{\rm W}},1}}-{e}^{-\Delta t/\tau_{{{\rm W}},2}}} \right )} &{\mbox{for}\ \ } 0\le \Delta t,
\end{array}\right .{\label{funcw}}
\end{eqnarray}
with
\begin{equation}
 \Delta t=t_{\rm post}-t_{\rm pre},
\end{equation}
where $t_{\rm{post}}{\mbox{ and }} t_{\rm{pre}}$ denote spike timing of presynaptic and postsynaptic neurons, respectively.
The asymmetric shape of the time window $W(\Delta t)$ is described in {{{Fig.}~\ref{figw}}}, where we set $\tau_{{{\rm W}},1}=25{\mbox{ [msec]}}$ and $\tau_{{{\rm W}},2}=2.5{\mbox{ [msec]}}$ as we set in what follows.

We encode the spatio-temporal patterns~$\tilde{s}_i^\mu$ according to this spike-timing-dependent synaptic plasticity.
In our previous study\cite{myosioka5}, we have introduced the learning rule 
\begin{equation}
 J_{ij}=\frac{1}{N}\sum_{\mu=1}^P\tilde{W}{\left ( {s_i^\mu-s_j^\mu} \right )},{\label{learningrule}}
\end{equation}
where, to take account of the periodicity of the present spatio-temporal patterns~$\tilde{s}_i^\mu$, we define $T$-periodic function $\tilde{W}{\left ( {\Delta t} \right )}$ of the form
\begin{eqnarray}
 \tilde{W}{\left ( {\Delta t} \right )}&=&\sum_{k=-\infty}^\infty W{\left ( {\Delta t +kT} \right )}\nonumber\\
&=&\frac{1}{\tau_{{{\rm W}},1}-\tau_{{{\rm W}},2}}{\left [ {\frac{e^{-\Delta t/\tau_{{{\rm W}},1}}-e^{-(T-\Delta t)/\tau_{{{\rm W}},1}}}{1-e^{-T/\tau_{{{\rm W}},1}}}-\frac{e^{-\Delta t/\tau_{{{\rm W}},2}}-e^{-(T-\Delta t)/\tau_{{{\rm W}},2}}}{1-e^{-T/\tau_{{{\rm W}},2}}}} \right ]},\nonumber\\
&& \qquad 0\le \Delta t<T.
\end{eqnarray}
This learning rule is applied also to the present neural networks.
As will be shown in the following sections, the spatio-temporal patterns encoded with this learning rule are retrieved successfully in the network dynamics~{{(\ref{vdyn})}}-{{(\ref{isum})}}.

\section{Perfect retrieval state}{\label{sect:stationarystate}}

Here we investigate the stationary properties of retrieval state of the network in the limit of infinite number of neurons.
In the present analysis, we focus on the retrieval state of the form
\begin{equation}
 t_i^\ast(k)=\frac{\tilde{T}}{Q}q_i^1+k\tilde{T},\qquad k=\ldots,-2,-1,0,1,2,\ldots,\ i=1,\ldots,N,{\label{perfectretrieval}}
\end{equation}
where we suppose pattern~1 as the retrieved pattern.
We term the retrieval state~{{(\ref{perfectretrieval})}} perfect retrieval state since no spike timing is allowed to deviate from the encoded pattern in this retrieval state.
Note that the period $\tilde{T}$ in {{{Eq.}~{(\ref{perfectretrieval})}}} is different from the period $T$, which is assumed in generating the spatio-temporal patterns~$\tilde{s}_i^\mu$, that is, the period of the retrieval process is different from the period of the encoded pattern.
In the present section, we aim to evaluate the period~$\tilde{T}$, which determines the form of the periodic solution for the perfect retrieval state.
The stability of the periodic solution is examined by a linear stability analysis in section~\ref{sect:stabilityanalysis}.

One possible way to determine the period~$\tilde{T}$ is substituting {{{Eq.}~{(\ref{perfectretrieval})}}} into {{{Eqs.}~{(\ref{ipp})}}} and {{(\ref{iip})}} so as to obtain the periodic synaptic electric current $I_i=I_{{{\rm PP}},i}+I_{{{\rm IP}}}$ in the limit of $N\rightarrow\infty$.
Then, the current $I_i$ is evaluated as a function of $\tilde{T}$, and hence we obtain the periodic firing pattern of $N$ neurons as a function of $\tilde{T}$.
Comparing the evaluated firing pattern with the substituted firing pattern~{{(\ref{perfectretrieval})}}, we can determine the period~$\tilde{T}$ self-consistently\cite{myosioka5}.

We follow the almost same scheme as above, although we make a slight detour for the convenience of the calculation below.
We first consider the solution of the form
\begin{equation}
 t_i(k)=t_q(k),\qquad k=\ldots,-2,-1,0,1,2,\ldots,\ i\in U_q,\ q=0,\ldots,Q-1,{\label{sublattice}}
\end{equation}
where the set of indexes $U_q$ is defined as
\begin{equation}
 U_q=\{i\ |\ q_i^1=q\}.
\end{equation}
We term a cluster of neurons that belong to $U_q$ sublattice~$q$.
In the solution~{{(\ref{sublattice})}}, neurons belonging to the same sublattice are assumed to behave in the same manner.
It will be shown that the dynamics~{{(\ref{vdyn})}}-{{(\ref{isum})}} has the solution of the form~{{(\ref{sublattice})}} in the limit of $N\rightarrow\infty$ if $P$ is finite\cite{yoshioka3}.
We will evaluate the $Q$-body dynamics for these sublattices, which has important implication for the stability analysis in section~\ref{sect:stabilityanalysis}.
After that, to this $Q$-body dynamics of sublattices, we substitute the solution of the form
\begin{equation}
 t_q^\ast(k)=\frac{\tilde{T}}{Q}q+k\tilde{T},\qquad k=\ldots,-2,-1,0,1,2,\ldots,\ q=0,\ldots,Q-1.{\label{retrievalstatesl}}
\end{equation}
Then, we obtain the period~$\tilde{T}$ for the perfect retrieval state~{{(\ref{perfectretrieval})}}.

In the analysis below, we always assume finite $P$ and finite $Q$.
Asterisks are used to indicate the variables in the stationary state.

\subsection{Dynamics of sublattices}

In order to evaluate the dynamics of $Q$ sublattices, we first evaluate the current $I_{{{\rm PP}},i}$ in the limit of $N\rightarrow\infty$ under the condition~{{(\ref{sublattice})}}.
Assuming that neuron~$i$ belongs to sublattice~$q$, we substitute {{{Eqs.}~{(\ref{learningrule})}}} and {{(\ref{sublattice})}} into {{{Eq.}~{(\ref{ipp})}}}.
Then, we have
\begin{eqnarray}
&&I_{{{\rm PP}},i}\nonumber\\
&=&A_{{{\rm PP}}}\sum_{q'=0}^{Q-1}\sum_{j\in U_{q'}} J_{ij}\sum_k S_{{{\rm PP}}}{\left [ {t-t_{q'}(k)} \right ]}\nonumber\\
&=&\frac{A_{{{\rm PP}}}}{Q}\sum_{q'}\frac{1}{N_{q'}}\sum_{j\in U_{q'}}\tilde{W}{\left [ {\frac{T}{Q}{\left ( {q-q'} \right )}} \right ]}\sum_k S_{{{\rm PP}}}{\left [ {t-t_{q'}(k)} \right ]}\nonumber\\
&&+\frac{A_{{{\rm PP}}}}{Q}\sum_{\mu>1}\sum_{q'}\frac{1}{N_{q'}}\sum_{j\in U_{q'}}\tilde{W}{\left [ {\frac{T}{Q}{\left ( {q_i^\mu-q_j^\mu} \right )}} \right ]}\sum_k S_{{{\rm PP}}}{\left [ {t-t_{q'}(k)} \right ]}\nonumber\\
&=&\frac{A_{{{\rm PP}}}}{Q}\sum_{q'}\tilde{J}_{q q'}\sum_k S_{{{\rm PP}}}{\left [ {t-t_{q'}(k)} \right ]},\qquad i\in U_q, {\label{ippslcal}}
\end{eqnarray}
where $N_q$ denotes the number of members in $U_q$, and variables $\tilde{J}_{qq'}$ and $\overline{W}$ are defined as
\begin{equation}
 \tilde{J}_{qq'}=\tilde{W}{\left [ {\frac{T}{Q}{\left ( {q-q'} \right )}} \right ]}+(P-1)\overline{W},{\label{newj}}
\end{equation}
\begin{equation}
 \overline{W}=\frac{1}{Q}\sum_{q=0}^{Q-1} \tilde{W}{\left ( {\frac{T}{Q}q} \right )}=\frac{1}{Q}\sum_{q=-\infty}^\infty W{\left ( {\frac{T}{Q}q} \right )}.{\label{avew}}
\end{equation}
{{{Equation}~{(\ref{ippslcal})}}} shows that the current~$I_{{{\rm PP}},i}$ is independent of index $i$ in the limit of $N\rightarrow\infty$.
Thus, we define the sublattice variable $I_{{{\rm PP}},q}$ as
\begin{equation}
 I_{{{\rm PP}},i}=I_{{{\rm PP}},q}=\frac{A_{{{\rm PP}}}}{Q}\sum_{q'}\tilde{J}_{q q'}\sum_k S_{{{\rm PP}}}{\left [ {t-t_{q'}(k)} \right ]},\qquad i\in U_{q},\ q=0,\ldots,Q-1.{\label{ippsl}}
\end{equation}
Following the same scheme, we rewrite the current~$I_{{{\rm IP}}}$ in {{{Eq.}~{(\ref{iip})}}} in the form
\begin{equation}
 I_{{{\rm IP}}}=\frac{A_{{{\rm IP}}}}{Q}\sum_{q'}\sum_k S_{{{\rm IP}}}{\left [ {t-t_{q'}(k)} \right ]}.{\label{iipsl}}
\end{equation}

Equations~{{(\ref{ippsl})}} and {{(\ref{iipsl})}} imply that the synaptic electric current $I_i=I_{{{\rm PP}},i}+I_{{{\rm IP}},i}$ depends only on $q$, that is, neurons belonging to the same sublattice obtain the same amount of synaptic electric current.
Therefore, the dynamics~{{(\ref{vdyn})}}-{{(\ref{isum})}} has the solution in which neurons belonging to the same sublattice behave in the same manner, as we have assumed in {{{Eq.}~{(\ref{sublattice})}}}.
Such dynamics of sublattices is expressed as
\begin{eqnarray}
\dot{v}_q&=&f{\left ( {v_q,w_{q1},\ldots,w_{qn}} \right )}+I_q,{\label{vdynsl}}\\
\dot{w}_{ql}&=&g_l{\left ( {v_q,w_{q1},\ldots,w_{qn}} \right )},\qquad l=1,\ldots,n,\ q=0,\ldots,Q-1{\label{wdynsl}}
\end{eqnarray}
with
\begin{equation}
 I_q=I_{{{\rm PP}},q}+I_{{{\rm IP}}}{\label{isumsl}}
\end{equation}
where ${\left ( {v_q,{\left\{ {w_{ql}} \right \}}} \right )}$ represents the common state of neurons that belong to sublattice~$q$.
The common synaptic electric current $I_{q}$ in {{{Eq.}~{(\ref{isumsl})}}} is defined by {{{Eqs.}~{(\ref{ippsl})}}} and {{(\ref{iipsl})}}.

\subsection{Derivation of perfect retrieval state}

Let us find the periodic solution for the perfect retrieval state~{{(\ref{retrievalstatesl})}} in the dynamics of sublattices {{(\ref{vdynsl})}}-{{(\ref{isumsl})}}.
Substituting {{{Eq.}~{(\ref{retrievalstatesl})}}} into {{{Eq.}~{(\ref{ippsl})}}}, we have
\begin{eqnarray}
I_{{{\rm PP}},q}^\ast&=&\frac{A_{{{\rm PP}}}}{Q}\sum_{q'}\tilde{J}_{qq'}\sum_k S_{{{\rm PP}}}{\left [ {t-{\left ( {\frac{\tilde{T}}{Q}q'+k\tilde{T}} \right )}} \right ]}\nonumber\\
&=&\frac{A_{{{\rm PP}}}}{Q}\sum_{q'}\tilde{J}_{qq'} \tilde{S}_{{{\rm PP}}}{\left ( {t-\frac{\tilde{T}}{Q}q'} \right )},{\label{ippast}}
\end{eqnarray}
where $\tilde{T}$-periodic function $\tilde{S}_{{{\rm PP}}}(t)$ is defined as
\begin{eqnarray}
 \tilde{S}_{{{\rm PP}}}(t)&=&\sum_k S_{{{\rm PP}}}{\left ( {t+k\tilde{T}} \right )}\nonumber\\
&=&\frac{1}{\tau_{{{\rm PP}},1}-\tau_{{{\rm PP}},2}}{\left [ {\frac{{e}^{-t/\tau_{{{\rm PP}},1}}}{1-{e}^{-\tilde{T}/\tau_{{{\rm PP}},1}}}-\frac{{e}^{-t/\tau_{{{\rm PP}},2}}}{1-{e}^{-\tilde{T}/\tau_{{{\rm PP}},2}}}} \right ]},\qquad 0\le t<\tilde{T}.
\end{eqnarray}
In the same manner, {{{Eq.}~{(\ref{iipsl})}}} is rewritten as
\begin{equation}
 I_{{{\rm IP}}}^\ast=\frac{A_{{{\rm IP}}}}{Q}\sum_{q'} \tilde{S}_{{{\rm IP}}}{\left ( {t-\frac{\tilde{T}}{Q}q'} \right )},{\label{iipast}}
\end{equation}
where $\tilde{T}$-periodic function $\tilde{S}_{{{\rm IP}}}(t)$ is defined as
\begin{eqnarray}
 \tilde{S}_{{{\rm IP}}}(t)&=&\sum_k S_{{{\rm IP}}}{\left ( {t+k\tilde{T}} \right )}\nonumber\\
&=&\frac{-1}{\tau_{{{\rm IP}},1}-\tau_{{{\rm IP}},2}}{\left [ {\frac{{e}^{-t/\tau_{{{\rm IP}},1}}}{1-{e}^{-\tilde{T}/\tau_{{{\rm IP}},1}}}-\frac{{e}^{-t/\tau_{{{\rm IP}},2}}}{1-{e}^{-\tilde{T}/\tau_{{{\rm IP}},2}}}} \right ]},\qquad 0\le t<\tilde{T}.
\end{eqnarray}
Therefore, $\tilde{T}$-periodic solution for the perfect retrieval state ${\left ( {v_q^\ast,{\left\{ {w_{ql}^\ast} \right \}}} \right )}$ obey the dynamics of the form
\begin{eqnarray}
\dot{v}^\ast_q&=&f{\left ( {v_q^\ast,w_{q1}^\ast,\ldots,w_{qn}^\ast} \right )}+I_q^\ast,{\label{vdynsingle}}\\
\dot{w}^\ast_{ql}&=&g_l{\left ( {v_q^\ast,w_{q1}^\ast,\ldots,w_{qn}^\ast} \right )},\qquad l=1,\ldots,n,\ q=0,\ldots,Q-1,{\label{wdynsingle}}
\end{eqnarray}
where
\begin{equation}
 I_q^\ast=I_{{{\rm PP}},q}^\ast+I_{{{\rm IP}}}^\ast.{\label{isumsingle}}
\end{equation}
As shown in {{{Eqs.}~{(\ref{ippast})}}} and {{(\ref{iipast})}}, $I_{{{\rm PP}},q}^\ast$ and $I_{{{\rm IP}}}^\ast$ are functions of $q$ and $\tilde{T}$, and also $I_q^\ast$ is a function of $q$ and $\tilde{T}$.
Hence, we can calculate the behavior of each sublattice as a function of $q$ and $\tilde{T}$ from the dynamics~{{(\ref{vdynsingle})}}-{{(\ref{isumsingle})}}.

Noting {{{Eqs.}~{(\ref{newj})}}}, {{(\ref{ippast})}}, and {{(\ref{iipast})}}, we obtain
\begin{equation}
 I_q^\ast{\left ( {t+\frac{\tilde{T}}{Q}q} \right )}= I_{q'}^\ast{\left ( {t+\frac{\tilde{T}}{Q}{q'}} \right )}.
\end{equation}
It means that every synaptic electric current~$I_{q}^\ast$ is identical, except that it exhibits the time shift according to $q$, and the behavior of all sublattices in {{{Eqs.}~{(\ref{vdynsingle})}}}-{{(\ref{isumsingle})}} are evaluated from the time shift of sublattice~0.
Therefore, we focus on the analysis of sublattice~0 in what follows.

We can calculate the behavior of sublattice~0 in the dynamics~{{(\ref{vdynsingle})}}-{{(\ref{isumsingle})}} for the arbitrary value of $\tilde{T}$.
In the stability analysis in section~\ref{sect:stabilityanalysis}, we will show that if the dynamics~{{(\ref{vdyn})}}-{{(\ref{isum})}} has the stable perfect retrieval state, then the dynamics~{{(\ref{vdynsingle})}}-{{(\ref{isumsingle})}} also has the stable periodic solution at $\tilde{T}=\tilde{T}^\ast$, where $\tilde{T}^\ast$ denotes the solution of the period now under consideration.
In almost every cases, for $\tilde{T}$ that is sufficiently close to $\tilde{T}^\ast$, sublattice~0 in the dynamics~{{(\ref{vdynsingle})}}-{{(\ref{isumsingle})}} exhibits the periodic firing motion, and hence the spike timing of sublattice~0 in the stationary state is written as\cite{myosioka5}
\begin{equation}
 t_0(k)=k\tilde{T}+r(\tilde{T}),\qquad k={\ldots,-2,-1,0,1,2\ldots}\label{spiketimingev}
\end{equation}
On the other hand, from {{{Eq.}~{(\ref{retrievalstatesl})}}}, we obtain the spike timing of sublattice~0 as
\begin{equation}
 t_0^\ast(k)=\frac{\tilde{T}^\ast}{Q}0+k\tilde{T}^\ast=k\tilde{T}^\ast,\qquad k={\ldots,-2,-1,0,1,2\ldots} \label{spiketimingas}
\end{equation}
Comparing {{{Eq.}~{(\ref{spiketimingev})}}} with {{{Eq.}~{(\ref{spiketimingas})}}}, we obtain the condition
\begin{equation}
 r(\tilde{T}^\ast)=0. {\label{conditionr}}
\end{equation}
As shown in the previous study\cite{myosioka5}, we can easily evaluate the explicit form of the function~$r(\tilde{T})$ numerically by integrating the single body dynamics of sublattice~0 in {{{Eqs.}~{(\ref{vdynsingle})}}}-{{(\ref{isumsingle})}}.
Once we evaluate the explicit form of the function $r(\tilde{T})$, we obtain the solution $\tilde{T}^\ast$ from the condition~{{(\ref{conditionr})}}. 

\subsection{Optimal form of the time window $W{\left ( {\Delta t} \right )}$ to encode multiple spatio-temporal patterns}

In general, the properties of the network depend on the number of stored patterns $P$.
We can encode a large number of patterns when the network exhibits the weak dependence on $P$.
To see to what extent the properties of the network depend on $P$, we decompose $I_q^\ast=I_{{{\rm PP}},q}^\ast+I_{{{\rm IP}}}^\ast$ defined by {{{Eqs.}~{(\ref{ippast})}}} and {{(\ref{iipast})}} into the form
\begin{equation}
 I_q^\ast=M_q^\ast+I_{{{\rm IP}}}^\ast+Z^\ast{\label{decompose}}
\end{equation}
with
\begin{equation}
M_q^\ast=\frac{A_{{{\rm PP}}}}{Q}\sum_{q'}\tilde{W}{\left [ {\frac{T}{Q}{\left ( {q-q'} \right )}} \right ]}\tilde{S}_{{{\rm PP}}}{\left ( {t-\frac{\tilde{T}}{Q}q'} \right )},
\end{equation}
\begin{equation}
Z^\ast=A_{{{\rm PP}}}(P-1)\overline{W}\ \overline{S}_{{{\rm PP}}}(t),{\label{crosstalk}}
\end{equation}
where
\begin{equation}
 \overline{S}_{{{\rm PP}}}(t)=\frac{1}{Q}\sum_{q'=0}^{Q-1}\tilde{S}_{{{\rm PP}}}{\left ( {t+\frac{\tilde{T}}{Q}q'} \right )}=\frac{1}{Q}\sum_{q'=-\infty}^{\infty}S_{{{\rm PP}}}{\left ( {t+\frac{\tilde{T}}{Q}q'} \right )}.
\end{equation}
We term $Z^\ast$ the crosstalk term since this term appears in {{{Eq.}~{(\ref{decompose})}}} as a result of encoding multiple spatio-temporal patterns.
The function $\overline{S}_{{{\rm PP}}}{\left ( {t} \right )}$ always takes the positive value.
Hence, as $P$ increases, $Z^\ast$ exhibits a increase or a decrease depending on the sign of $\overline{W}$, until the perfect retrieval state breaks at the critical number of patterns $P=P^c$.

Let us take notice of $\overline{W}$ appearing in {{{Eq.}~{(\ref{crosstalk})}}}.
The quantity $\overline{W}$, which is defined by {{{Eq.}~{(\ref{avew})}}}, is the average of the time window $W{\left ( {\Delta t} \right )}$.
For the time window $W{\left ( {\Delta t} \right )}$ defined by {{{Eq.}~{(\ref{funcw})}}}, one can easily show
\begin{equation}
 \overline{W}=0.{\label{vanish}}
\end{equation}
In this case, the crosstalk term $Z^\ast$ vanishes, and hence we can encode the arbitrary number of patterns in the limit of $N\rightarrow\infty$ as far as $P$ is finite.
It turns out that the present form of the time window $W{\left ( {\Delta t} \right )}$, which is  found in experiments, is of great advantage to reduce the size of $\overline{W}$ and also the crosstalk among encoded patterns.

\section{Stability of the perfect retrieval state}{\label{sect:stabilityanalysis}}

Although we have derived the periodic solutions for the perfect retrieval state in the previous section, it still remains unclear whether the derived periodic solutions are stable in the network dynamics~{{(\ref{vdyn})}}-{{(\ref{isum})}}.
In some cases, the derivation of periodic solutions in the previous section yields unstable solutions, and the network cannot settle into such unstable retrieval state.
In the present section, we employ a linear stability analysis for the perfect retrieval state we have derived in the previous section.
That is the application of Floquet theory, which yields an eigenvalue problem with the finite size of the matrix.

\subsection{Decomposition of the problem: stability of sublattices and stability of the perfect retrieval state in the dynamics of sublattices~{{(\ref{vdynsingle})}}-{{(\ref{isumsingle})}}}

In a linear stability analysis, infinitesimal perturbation is assumed in the initial condition, and then the time evolution of the deviation from the target solution is investigated to the first order in Taylor series expansion.
When we apply Floquet theory to the present system, the spike timing of neuron~$i$ that belongs to sublattice~$q$ is written in the form 
\begin{equation}
 t_i(k)=t_q^\ast(k)+\delta t_i(k), \qquad k={\ldots,-2,-1,0,1,2\ldots},\ i\in U_q,\ q=0,\ldots,Q-1,{\label{deltat}}
\end{equation}
where we suppose pattern~1 as the retrieved pattern.
We assume that the initial condition is correlated only with pattern~1 and the correlation with other patterns does not arise in the time evolution of the network dynamics, that is, we assume $\delta t_i(k)$ is correlated only with $s_j^1\ (j=1,\ldots,N)$.
Substituting {{{Eqs.}~{(\ref{learningrule})}}} and {{(\ref{deltat})}} into {{{Eq.}~{(\ref{ipp})}}}, we obtain $I_{{{\rm PP}},i}\ (i\in U_q)$ of the form
\begin{eqnarray}
&&I_{{{\rm PP}},i}\nonumber\\
&=&\frac{A_{{{\rm PP}}}}{Q}\sum_{q'}\tilde{W}{\left [ {\frac{T}{Q}{\left ( {q-q'} \right )}} \right ]}\sum_k \frac{1}{N_{q'}}\sum_{j\in U_{q'}}S_{{{\rm PP}}}{\left [ {t-t_{q'}^\ast(k)-\delta t_j(k)} \right ]}\nonumber\\
&&+\frac{A_{{{\rm PP}}}}{Q}\sum_{q'}\sum_k\sum_{\mu>1}\frac{1}{N_{q'}}\sum_{j\in U_{q'}}\tilde{W}{\left [ {\frac{T}{Q}{\left ( {q_i^\mu-q_j^\mu} \right )}} \right ]} S_{{{\rm PP}}}{\left [ {t-t_{q'}^\ast(k)-\delta t_j(k)} \right ]}\nonumber\\
&=&\frac{A_{{{\rm PP}}}}{Q}\sum_{q'}\tilde{W}{\left [ {\frac{T}{Q}{\left ( {q-q'} \right )}} \right ]}\sum_k\frac{1}{N_{q'}}\sum_{j\in U_{q'}} S_{{{\rm PP}}}{\left [ {t-t_{q'}^\ast(k)-\delta t_j(k)} \right ]}\nonumber\\
&&+\frac{A_{{{\rm PP}}}}{Q}\sum_{q'}\sum_k\sum_{\mu>1}{\left ( {\frac{1}{N_{q'}}\sum_{j\in U_{q'}}\tilde{W}{\left [ {\frac{T}{Q}{\left ( {q_i^\mu-q_j^\mu} \right )}} \right ]}} \right )}{\left ( {\frac{1}{N_{q'}}\sum_{j\in U_{q'}} S_{{{\rm PP}}}{\left [ {t-t_{q'}^\ast(k)-\delta t_j(k)} \right ]}} \right )}\nonumber\\
&=&\frac{A_{{{\rm PP}}}}{Q}\sum_{q'}\tilde{J}_{qq'}\sum_k \frac{1}{N_{q'}}\sum_{j\in U_{q'}}  S_{{{\rm PP}}}{\left [ {t-t_{q'}^\ast(k)-\delta t_j(k)} \right ]},\qquad i\in U_q, {\label{introjdelta}}
\end{eqnarray}
where we utilize the assumption that $\delta t_i(k)$ is correlated only with $s_j^1=\frac{T}{Q}q_j^1\ (j=1,\ldots,N).$
Since {{{Eq.}~{(\ref{introjdelta})}}} shows that $I_{{{\rm PP}},i}$ depends only on $q$, we are allowed to define sublattice variable $I_{{{\rm PP}},q}$ as
\begin{equation}
 I_{{{\rm PP}},i}=I_{{{\rm PP}},q}=\frac{A_{{{\rm PP}}}}{Q}\sum_{q'}\tilde{J}_{qq'}\sum_k \frac{1}{N_{q'}}\sum_{j\in U_{q'}}  S_{{{\rm PP}}}{\left [ {t-t_{q'}^\ast(k)-\delta t_j(k)} \right ]},\qquad i\in U_q.{\label{ippq}}
\end{equation}
Performing a truncated Taylor series expansion of {{{Eq.}~{(\ref{ippq})}}}, we have
\begin{equation}
 I_{{{\rm PP}},q}=I_{{{\rm PP}},q}^\ast+\delta I_{{{\rm PP}},q}
\end{equation}
with
\begin{equation}
\delta I_{{{\rm PP}},q}=-\frac{A_{{{\rm PP}}}}{Q}\sum_{q'}\tilde{J}_{qq'}\sum_k S_{{{\rm PP}}}'{\left [ {t-t_{q'}^\ast(k)} \right ]}\delta\overline{t}_{q'}(k),{\label{ippdelta}}
\end{equation}
where the derivative of $S_{{{\rm PP}}}(t)$ is written as
\begin{equation}
S'_{{{\rm PP}}}(t)=\left\{\begin{array}{lc}
0 &{\mbox{for}\ \ } t<0,\\
\displaystyle \frac{-1}{\tau_{{{\rm PP}},1}-\tau_{{{\rm PP}},2}}{\left ( {\frac{1}{\tau_{{{\rm PP}},1}}{e}^{-t/\tau_{{{\rm PP}},1}}-\frac{1}{\tau_{{{\rm PP}},2}}{e}^{-t/\tau_{{{\rm PP}},2}}} \right )}&{\mbox{for}\ \ } 0\le t,
\end{array}\right .
\end{equation}
and the sublattice variable $\delta\overline{t}_q(k)$ is defined as
\begin{equation}
  \delta\overline{t}_q(k)=\frac{1}{N_q}\sum_{i\in U_q}\delta t_i(k).{\label{meantime}}
\end{equation}

Following the same scheme as $I_{{{\rm PP}},q}$, we obtain the deviation of $I_{{{\rm IP}}}$ in {{{Eq.}~{(\ref{iip})}}} as
\begin{equation}
I_{{{\rm IP}}}=I_{{{\rm IP}}}^\ast+\delta I_{{{\rm IP}}},
\end{equation}
with
\begin{equation}
 \delta I_{{{\rm IP}}}=-\frac{A_{{{\rm IP}}}}{Q}\sum_{q'}\sum_k S_{{{\rm IP}}}'{\left [ {t-t_{q'}(k)} \right ]}\delta \overline{t}_{q'}(k),{\label{iipdelta}}
\end{equation}
where the derivative of $S_{{{\rm IP}}}(t)$ is written as
\begin{equation}
S'_{{{\rm IP}}}(t)=\left\{\begin{array}{lc}
0 &{\mbox{for}\ \ } t<0,\\
\displaystyle \frac{1}{\tau_{{{\rm IP}},1}-\tau_{{{\rm IP}},2}}{\left ( {\frac{1}{\tau_{{{\rm IP}},1}}{e}^{-t/\tau_{{{\rm IP}},1}}-\frac{1}{\tau_{{{\rm IP}},2}}{e}^{-t/\tau_{{{\rm IP}},2}}} \right )}&{\mbox{for}\ \ } 0\le t.
\end{array}\right . \\
\end{equation}

We represent deviation appearing in the state of neuron~$i$ by 
\begin{eqnarray}
v_i&=&v_q^\ast+\delta v_i,\\
w_{il}&=&w_{ql}^\ast+\delta w_{il},\qquad l=1,\ldots,n,\ i\in U_q,\ q=0,\ldots,Q-1.
\end{eqnarray}
Noting {{{Eq.}~{(\ref{ippq})}}}, we safely replace $I_i=I_{{{\rm PP}},i}+I_{{{\rm IP}}}$ in {{{Eq.}~{(\ref{vdyn})}}} by sublattice variable $I_q=I_{{{\rm PP}},q}+I_{{{\rm IP}}}$.
Then, we perform a truncated Taylor series expansion of {{{Eqs.}~{(\ref{vdyn})}}}-{{(\ref{wdyn})}} and obtain the dynamics of the form
\begin{eqnarray}
\delta \dot{v}_i&=&{\left .\frac{\partial {f}}{\partial {v}} \right |_{{q}}}\delta v_i+\sum_{l'} {\left .\frac{\partial {f}}{\partial {w_{l'}}} \right |_{{q}}}\delta w_{il'} +\delta I_q,{\label{vdyndelta}}\\
\delta \dot{w}_{il}&=&{\left .\frac{\partial {g_l}}{\partial {v}} \right |_{{q}}}\delta v_i+\sum_{l'} {\left .\frac{\partial {g_l}}{\partial {w_{l'}}} \right |_{{q}}}\delta w_{il'},\qquad l=1,\ldots,n,\ i\in U_q,\ q=0,\ldots,Q-1{\label{wdyndelta}}
\end{eqnarray}
with
\begin{equation}
\delta I_q=\delta I_{{{\rm PP}},q}+\delta I_{{{\rm IP}}},{\label{isumdelta}}
\end{equation}
where we introduce abbreviations such as ${\left .\frac{\partial {f}}{\partial {v}} \right |_{{q}}}={\left .\frac{\partial {f}}{\partial {v}} \right |_{{{\left ( {v_q^\ast,{\left\{ {w_{ql}^\ast} \right \}}} \right )}}}}$.

From the definition of spike timing, we have  $v_i{\left [ {t_q^\ast(k)+\delta t_i(k)} \right ]}=\theta\ (i\in U_q)$, which yields
\begin{equation}
 \delta t_i(k)=-\frac{\delta v_i{\left [ {t_q^\ast(k)} \right ]}}{c},\qquad i\in U_q,{\label{respiketiming}}
\end{equation}
where constant $c$ is defined as
\begin{equation}
 c=\dot{v}_q^\ast{{\left [ {t_{{q}}^\ast({k})} \right ]}}.
\end{equation}
Note that constant $c$ is independent of $q{\mbox{ and }} k.$
Now we can evaluate the time evolution of ${\left ( {\delta v_i,{\left\{ {\delta w_{il}} \right \}}} \right )}$ from {{{Eqs.}~{(\ref{vdyndelta})}}}-{{(\ref{respiketiming})}}.
To solve this dynamics we need to calculate $\delta I_q=\delta I_{{{\rm PP}},q}+\delta  I_{{{\rm IP}}}$, in which $\delta \overline{t}_q(k)$ are required at time $t=t_q^\ast(k)\ (k={\ldots,-2,-1,0,1,2\ldots})$ as shown in {{{Eqs.}~{(\ref{ippdelta})}}} and {{(\ref{iipdelta})}}.
We can evaluate $\delta \overline{t}_q(k)$ from ${\left\{ {\delta v_i{\left [ {t_q^\ast(k)} \right ]}} \right \}}$ by use of {{{Eqs.}~{(\ref{meantime})}}} and {{(\ref{respiketiming})}}.

It is a hopeless task to apply Floquet theory directly to the $N$-body dynamics~{{(\ref{vdyndelta})}}-{{(\ref{isumdelta})}} since that gives the eigenvalue problem with the infinite size of matrix.
For the purpose of reducing the degree of freedom, we define the following sublattice variables
\begin{eqnarray}
\overline{v}_q&=&\frac{1}{N_q}\sum_{i\in U_q} v_i,\\
\overline{w}_{ql}&=&\frac{1}{N_q}\sum_{i\in U_q} w_{il},\qquad l=1,\ldots,n,\ q=0,\ldots,Q-1.
\end{eqnarray}
Then, from {{{Eqs.}~{(\ref{vdyndelta})}}} and {{(\ref{wdyndelta})}}, we have
\begin{eqnarray}
\delta \dot{\overline{v}}_q&=&\frac{1}{N_q}\sum_{i\in U_q}\delta \dot{v}_i={\left .\frac{\partial {f}}{\partial {v}} \right |_{{q}}}\delta\overline{v}_q+\sum_{l'}{\left .\frac{\partial {f}}{\partial {w_{l'}}} \right |_{{q}}}\delta\overline{w}_{ql'}+\delta I_q,{\label{vdynq}}\\
 \delta \dot{\overline{w}}_{ql}&=&\frac{1}{N_q}\sum_{i\in U_q} \delta \dot{w}_{il}={\left .\frac{\partial {g_l}}{\partial {v}} \right |_{{q}}}\delta\overline{v}_q+\sum_{l'}{\left .\frac{\partial {g_l}}{\partial {w_{l'}}} \right |_{{q}}}\delta\overline{w}_{ql'},\nonumber\\
&&\qquad l=1,\ldots,n,\ q=0,\ldots,Q-1,{\label{wdynq}}
\end{eqnarray}
where, from {{{Eqs.}~{(\ref{ippdelta})}}} and {{(\ref{iipdelta})}}, $\delta I_q$ in {{{Eq.}~{(\ref{vdynq})}}} is written as
\begin{eqnarray}
 \delta I_q&=&\delta I_{{{\rm PP}},q}+\delta I_{{{\rm IP}}}\nonumber\\
&=&-\frac{A_{{{\rm PP}}}}{Q}\sum_{q'}\tilde{J}_{qq'}\sum_k S_{{{\rm PP}}}'{\left [ {t-t_{q'}^\ast(k)} \right ]}\delta\overline{t}_{q'}(k)-\frac{A_{{{\rm IP}}}}{Q}\sum_{q'}\sum_k S_{{{\rm IP}}}'{\left [ {t-t_{q'}(k)} \right ]}\delta \overline{t}_{q'}(k).
\end{eqnarray}
In addition, substituting {{{Eq.}~{(\ref{respiketiming})}}} into {{{Eq.}~{(\ref{meantime})}}}, we have
\begin{equation}
 \delta \overline{t}_q(k)=-\frac{\delta \overline{v}_q{\left [ {t_q^\ast(k)} \right ]}}{c}.{\label{rerespiketiming}}
\end{equation}
Now we obtain the $Q$-body dynamics~{{(\ref{vdynq})}}-{{(\ref{rerespiketiming})}}.
Calculation of $\delta I_q$ requires ${\left\{ {\delta \overline{t}_{q'}(k)} \right \}}$, which are obtained from ${\left\{ {\delta\overline{v}_{q'}{\left [ {t_{q'}^\ast(k)} \right ]}} \right \}}$ together with {{{Eq.}~{(\ref{rerespiketiming})}}}.
To this $Q$-body dynamics we will apply Floquet theory in section~\ref{sect:applicationfloquet}.

The stability of the periodic solution in the $Q$-body dynamics~{{(\ref{vdynq})}}-{{(\ref{rerespiketiming})}} is the necessary condition for the stability of the retrieval state in the original dynamics~{{(\ref{vdyn})}}-{{(\ref{isum})}}, but not the sufficient condition.
Therefore, we must investigate the behavior of the following variables
\begin{eqnarray}
\delta\tilde{v}_i&=&v_i-\overline{v}_q=\delta v_i-\delta \overline{v}_q,\\
\delta\tilde{w}_{il}&=&w_{il}-\overline{w}_{ql}=\delta w_{il}-\delta \overline{w}_{ql},\qquad l=1,\ldots,n,\ i\in U_q,\ q=0,\ldots,Q-1.
\end{eqnarray}
If the perfect retrieval state is stable, ${\left ( {\delta\tilde{v}_i,{\left\{ {\delta\tilde{w}_{il} } \right \}}} \right )}$ converges into the fixed point ${\left ( {0,{\left\{ {0} \right \}}} \right )}$.
Subtracting {{{Eqs.}~{(\ref{vdynq})}}} and {{(\ref{wdynq})}} from {{{Eqs.}~{(\ref{vdyndelta})}}} and {{(\ref{wdyndelta})}} respectively, we obtain
\begin{eqnarray}
\delta\dot{\tilde{v}}_i&=&{\left .\frac{\partial {f}}{\partial {v}} \right |_{{q}}}\delta\tilde{v}_i+\sum_{l'}{\left .\frac{\partial {f}}{\partial {w_{l'}}} \right |_{{q}}}\delta\tilde{w}_{il'},{\label{vdyndeltasingle}}\\
 \delta\dot{\tilde{w}}_{il}&=&{\left .\frac{\partial {g_l}}{\partial {v}} \right |_{{q}}}\delta\tilde{v}_i+\sum_{l'}{\left .\frac{\partial {g_l}}{\partial {w_{l'}}} \right |_{{q}}}\delta\tilde{w}_{il'},\qquad l=0,\ldots,n,\ i\in U_q,\ q=0,\ldots,Q-1.{\label{wdyndeltasingle}}
\end{eqnarray}
For the stable perfect retrieval state, the fixed point ${\left ( {0,{\left\{ {0} \right \}}} \right )}$ is necessary to be stable in the dynamics~{{(\ref{vdyndeltasingle})}} and {{(\ref{wdyndeltasingle})}}.
Note that $N$ deviations ${\left ( {\delta\tilde{v}_i,{\left\{ {\delta\tilde{w}_{il} } \right \}}} \right )}$  appearing in the dynamics~{{{Eqs.}~{(\ref{vdyndeltasingle})}}} and {{(\ref{wdyndeltasingle})}} do not interact with each other since this dynamics includes no interaction term like $\delta I_q$.
This stability problem is thus a single body problem, which is easily evaluated numerically.

The stability problem of the perfect retrieval state in the dynamics~{{(\ref{vdyn})}}-{{(\ref{isum})}} is now decomposed into two stability problems: the stability of the perfect retrieval state in the $Q$-body dynamics~{{(\ref{vdynq})}}-{{(\ref{rerespiketiming})}} and the stability of the fixed point ${\left ( {0,{\left\{ {0} \right \}}} \right )}$ in the single-body dynamics~{{(\ref{vdyndeltasingle})}} and {{(\ref{wdyndeltasingle})}}.
What are the implications of these two stability problems?
It is straightforward to see that the former problem is equivalent to the stability problem of the perfect retrieval state in the $Q$-body dynamics of sublattices~{{(\ref{vdynsl})}}-{{(\ref{isumsl})}}.
Hence, we conveniently call the former problem the stability of the perfect retrieval state in the dynamics of sublattices.
In the dynamics of sublattices~{{(\ref{vdynsl})}}-{{(\ref{isumsl})}} we neglect a distribution of spike timing of neurons in each sublattice, and this distribution of spike timing is treated in the latter problem.
We thus term the latter problem the stability of sublattices.

It is of interest that a truncated Taylor series expansion of {{{Eqs.}~{(\ref{vdynsingle})}}} and {{(\ref{wdynsingle})}} with fixed $I_q^\ast$ gives the same stability problem as {{{Eqs.}~{(\ref{vdyndeltasingle})}}} and {{(\ref{wdyndeltasingle})}}.
This result implies that if the periodic solution ${\left ( {v_q^\ast,{\left\{ {w_{ql}^\ast} \right \}}} \right )}$ is stable in the dynamics ~{{(\ref{vdynsingle})}}-{{(\ref{isumsingle})}}, then the stability of sublattices are ensured.
We evaluate the periodic solution ${\left ( {v_q^\ast,{\left\{ {w_{ql}^\ast} \right \}}} \right )}$ by the numerical integration of the dynamics~{{(\ref{vdynsingle})}} and {{(\ref{wdynsingle})}}, and hence it is impossible to obtain the unstable periodic solution of the dynamics~{{(\ref{vdynsingle})}} and {{(\ref{wdynsingle})}}.
In other words, the numerically evaluated periodic solution~${\left ( {v_q^\ast,{\left\{ {w_{ql}^\ast} \right \}}} \right )}$ is always stable, and also the stability of sublattices is always ensured.
Therefore, further investigation on the stability of sublattice is unnecessary, and we focus on the stability of the perfect retrieval state in the dynamics of sublattices in the next section.

\subsection{Floquet theory for the perfect retrieval state in the dynamics of sublattices}{\label{sect:applicationfloquet}}

Here we apply Floquet theory to the $Q$-body dynamics~{{(\ref{vdynq})}}-{{(\ref{rerespiketiming})}}.
In the evaluation of this dynamics, $\delta\overline{t}_{q}(k)$ are required at time $t=t_q^\ast(k)\ (k={\ldots,-2,-1,0,1,2\ldots})$.
One may thus consider it convenient to define the vector 
\begin{equation}
 \delta {\bf x}_q(k)={\left ( {\delta \overline{v}_q{{\left [ {t_{{q}}^\ast({k})} \right ]}},\delta \overline{w}_{q1}{{\left [ {t_{{q}}^\ast({k})} \right ]}},\ldots,\delta \overline{w}_{qn}{{\left [ {t_{{q}}^\ast({k})} \right ]}}} \right )}.
\end{equation}
The vector $\delta {\bf x}_q(k)$ represents the deviation at time $t=t_q^\ast(k)$.
Since the vector $\delta {\bf x}_q(k)$ includes the variable $\delta\overline{v}_q{{\left [ {t_{{q}}^\ast({k})} \right ]}}$, we can calculate $\delta\overline{t}_q(k)$ from $\delta {\bf x}_q(k)$ by use of {{{Eq.}~{(\ref{rerespiketiming})}}}.
Let us consider the problem of calculating the deviation $\delta {\bf x}_0(k+1)$ from the past deviations $\delta {\bf x}_q(k')\ (q=0,\ldots,Q-1,\ k'<k+1)$.

The $\alpha$-functions $S_{{{\rm PP}}}(t)$ and $S_{{{\rm IP}}}(t)$ give an infinite long-time influence after the activation, and the derivatives of these $\alpha$-functions appearing in {{{Eqs.}~{(\ref{ippdelta})}}} and {{(\ref{iipdelta})}} also have an infinite long-time influence.
It means that long past deviations $\delta\overline{t}_q(k')$ and also $\delta{\bf x}_q(k')$ are necessary in the evaluation of the present value of $\delta I_0.$
It is again a hopeless task to consider Floquet theory based on the vector $\delta{\bf x}_q(k)$ since that still gives an eigenvalue problem with the infinite size of matrix.

For the further reduction of the size of matrix, we define the variables
\begin{eqnarray}
I_{1,q}&=&\frac{A_{{{\rm PP}}}}{Q}\sum_{q'}\tilde{J}_{qq'}\sum_{t_{q'}^\ast(k')<t}\frac{{e}^{-{\left [ {t-t_{q'}^\ast(k')-\delta\overline{t}_{q'}(k')} \right ]}/\tau_{{{\rm PP}},1}}}{\tau_{{{\rm PP}},1}-\tau_{{{\rm PP}},2}},{\label{i1}}\\
I_{2,q}&=&\frac{A_{{{\rm PP}}}}{Q}\sum_{q'}\tilde{J}_{qq'}\sum_{t_{q'}^\ast(k')<t}\frac{-{e}^{-{\left [ {t-t_{q'}^\ast(k')-\delta\overline{t}_{q'}(k')} \right ]}/\tau_{{{\rm PP}},2}}}{\tau_{{{\rm PP}},1}-\tau_{{{\rm PP}},2}},\\
I_{3,q}&=&\frac{A_{{{\rm IP}}}}{Q}\sum_{q'}\sum_{t_{q'}^\ast(k')<t}\frac{-{e}^{-{\left [ {t-t_{q'}^\ast(k')-\delta\overline{t}_{q'}(k')} \right ]}/\tau_{{{\rm IP}},1}}}{\tau_{{{\rm IP}},1}-\tau_{{{\rm IP}},2}},\\
I_{4,q}&=&\frac{A_{{{\rm IP}}}}{Q}\sum_{q'}\sum_{t_{q'}^\ast(k')<t}\frac{{e}^{-{\left [ {t-t_{q'}^\ast(k')-\delta\overline{t}_{q'}(k')} \right ]}/\tau_{{{\rm IP}},2}}}{\tau_{{{\rm IP}},1}-\tau_{{{\rm IP}},2}}.{\label{i4}}
\end{eqnarray}
Then, for the specific form of $\alpha$-function~{{(\ref{spp})}}, we can rewrite {{{Eq.}~{(\ref{ippdelta})}}} as
\begin{eqnarray}
 \delta I_{{{\rm PP}},0}&=&-\frac{A_{{{\rm PP}}}}{Q}\sum_{q'}\tilde{J}_{0q'}S_{{{\rm PP}}}'{\left [ {t-t_{q'}^\ast(k)} \right ]}\delta\overline{t}_{q'}(k)\nonumber\\
&&+\delta I_{1,0}{{\left [ {t_{{0}}^\ast({k})} \right ]}}{e}^{-{\left [ {t-t_0^\ast(k)} \right ]}/\tau_{{{\rm PP}},1}}+\delta I_{2,0}{{\left [ {t_{{0}}^\ast({k})} \right ]}}{e}^{-{\left [ {t-t_0^\ast(k)} \right ]}/\tau_{{{\rm PP}},2}},\nonumber\\
&&\qquad t_0^\ast(k)<t<t_0^\ast(k+1),{\label{ippreduce}}
\end{eqnarray}
where
\begin{eqnarray}
\delta I_{1,0}{{\left [ {t_{{0}}^\ast({k})} \right ]}}&=&-\frac{A_{{{\rm PP}}}}{Q}\sum_{q'}\tilde{J}_{0q'}\sum_{t_{q'}^\ast(k')<t_0^\ast(k)}\frac{-{e}^{-{\left [ {t_0^\ast(k)-t_{q'}^\ast(k')} \right ]}/\tau_{{{\rm PP}},1}}}{\tau_{{{\rm PP}},1}{\left ( {\tau_{{{\rm PP}},1}-\tau_{{{\rm PP}},2}} \right )}}\delta \overline{t}_{q'}(k'),{\label{i10delta}}\\
\delta I_{2,0}{{\left [ {t_{{0}}^\ast({k})} \right ]}}&=&-\frac{A_{{{\rm PP}}}}{Q}\sum_{q'}\tilde{J}_{0q'}\sum_{t_{q'}^\ast(k')<t_0^\ast(k)}\frac{{e}^{-{\left [ {t_0^\ast(k)-t_{q'}^\ast(k')} \right ]}/\tau_{{{\rm PP}},2}}}{\tau_{{{\rm PP}},2}{\left ( {\tau_{{{\rm PP}},1}-\tau_{{{\rm PP}},2}} \right )}}\delta \overline{t}_{q'}(k').
\end{eqnarray}
In the same way, we rewrite {{{Eq.}~{(\ref{iipdelta})}}} as
\begin{eqnarray}
 \delta I_{{{\rm IP}}}&=&-\frac{A_{{{\rm PP}}}}{Q}\sum_{q'}S_{{{\rm IP}}}'{\left [ {t-t_{q'}^\ast(k)} \right ]}\delta\overline{t}_{q'}(k)\nonumber\\
&&+\delta I_{3,0}{{\left [ {t_{{0}}^\ast({k})} \right ]}}{e}^{-{\left [ {t-t_0^\ast(k)} \right ]}/\tau_{{{\rm IP}},1}}+\delta I_{4,0}{{\left [ {t_{{0}}^\ast({k})} \right ]}}{e}^{-{\left [ {t-t_0^\ast(k)} \right ]}/\tau_{{{\rm IP}},2}},\nonumber\\
&&\qquad t_0^\ast(k)<t<t_0^\ast(k+1),{\label{iipreduce}}
\end{eqnarray}
where
\begin{eqnarray}
\delta I_{3,0}{{\left [ {t_{{0}}^\ast({k})} \right ]}}&=&-\frac{A_{{{\rm IP}}}}{Q}\sum_{q'}\sum_{t_{q'}^\ast(k')<t_0^\ast(k)}\frac{{e}^{-{\left [ {t_0^\ast(k)-t_{q'}^\ast(k')} \right ]}/\tau_{{{\rm IP}},1}}}{\tau_{{{\rm IP}},1}{\left ( {\tau_{{{\rm IP}},1}-\tau_{{{\rm IP}},2}} \right )}}\delta \overline{t}_{q'}(k'),\\
\delta I_{4,0}{{\left [ {t_{{0}}^\ast({k})} \right ]}}&=&-\frac{A_{{{\rm IP}}}}{Q}\sum_{q'}\sum_{t_{q'}^\ast(k')<t_0^\ast(k)}\frac{-{e}^{-{\left [ {t_0^\ast(k)-t_{q'}^\ast(k')} \right ]}/\tau_{{{\rm IP}},2}}}{\tau_{{{\rm IP}},2}{\left ( {\tau_{{{\rm IP}},1}-\tau_{{{\rm IP}},2}} \right )}}\delta \overline{t}_{q'}(k').
\end{eqnarray}
In {{{Eqs.}~{(\ref{ippreduce})}}} and {{(\ref{iipreduce})}}, $\delta I_{{{\rm PP}},0}$ and $\delta I_{{{\rm IP}}}$\  $(t_0^\ast(k)<t<t_0^\ast(k+1))$ are evaluated only from ${\left\{ {\delta\overline{t}_{q'}(k)} \right \}}$ and ${\left\{ {\delta I_{s',0}{{\left [ {t_{{0}}^\ast({k})} \right ]}}} \right \}}$.
Solving the dynamics~{{(\ref{vdynq})}} and {{(\ref{wdynq})}} with $\delta I_0=\delta I_{{{\rm PP}},0}+\delta I_{{{\rm IP}}}$ defined by {{{Eqs.}~{(\ref{ippreduce})}}} and {{(\ref{iipreduce})}} under the condition ${\left ( {\delta\overline{v}_0{{\left [ {t_{{0}}^\ast({k})} \right ]}}, {\left\{ {\delta\overline{w}_{0l'}{{\left [ {t_{{0}}^\ast({k})} \right ]}}} \right \}}} \right )}$, we obtain the next deviation ${\left ( {\delta\overline{v}_0{{\left [ {t_{{0}}^\ast({k+1})} \right ]}},{\left\{ {\delta\overline{w}_{0l'}{{\left [ {t_{{0}}^\ast({k+1})} \right ]}}} \right \}}} \right )}$ as a function of ${\left\{ {\delta\overline{t}_{q'}(k)} \right \}}$, $\delta\overline{v}_0{{\left [ {t_{{0}}^\ast({k})} \right ]}}$, ${\left\{ {\delta\overline{w}_{0l'}{{\left [ {t_{{0}}^\ast({k})} \right ]}}} \right \}}$, and ${\left\{ {\delta I_{s',0}{{\left [ {t_{{0}}^\ast({k})} \right ]}}} \right \}}$.
Hence, we are allowed to define the functions
\begin{eqnarray}
\delta v_0{{\left [ {t_{{0}}^\ast({k+1})} \right ]}}&=&R{\left ( {{\left\{ {\delta\overline{t}_{q'}(k)} \right \}},\delta\overline{v}_0{{\left [ {t_{{0}}^\ast({k})} \right ]}},{\left\{ {\delta\overline{w}_{0l'}{{\left [ {t_{{0}}^\ast({k})} \right ]}}} \right \}},{\left\{ {\delta I_{s',0}{{\left [ {t_{{0}}^\ast({k})} \right ]}}} \right \}}} \right )},{\label{defrf}}\\
\delta w_{0l}{{\left [ {t_{{0}}^\ast({k+1})} \right ]}}&=&S_l{\left ( {{\left\{ {\delta\overline{t}_{q'}(k)} \right \}},\delta\overline{v}_0{{\left [ {t_{{0}}^\ast({k})} \right ]}},{\left\{ {\delta\overline{w}_{0l'}{{\left [ {t_{{0}}^\ast({k})} \right ]}}} \right \}},{\left\{ {\delta I_{s',0}{{\left [ {t_{{0}}^\ast({k})} \right ]}}} \right \}}} \right )},\nonumber\\
&&\qquad l=1,\ldots,n.{\label{defsf}}
\end{eqnarray}
Because of the form of the dynamics~{{(\ref{vdynq})}} and {{(\ref{wdynq})}}, we obtain
\begin{eqnarray}
&&R{\left ( {{\left\{ {\delta\overline{t}_{q'}} \right \}},\delta\overline{v},{\left\{ {\delta\overline{w}_{l'}} \right \}},{\left\{ {\delta I_{s'}} \right \}}} \right )}\nonumber\\
&=&\sum_{q'}{\frac{\partial {R}}{\partial{\left ( {{\delta\overline{t}_{q'}}} \right )}}}\delta\overline{t}_{q'}+{\frac{\partial {R}}{\partial{\left ( {{\delta\overline{v}}} \right )}}}\delta\overline{v}+\sum_{l'}{\frac{\partial {R}}{\partial{\left ( {{\delta\overline{w}_{l'}}} \right )}}}\delta\overline{w}_{l'}+\sum_{s'}{\frac{\partial {R}}{\partial{\left ( {{\delta{I}_{s'}}} \right )}}}\delta{I}_{s'},{\label{rfc}}\\
&&S_l{\left ( {{\left\{ {\delta\overline{t}_{q'}} \right \}},\delta\overline{v},{\left\{ {\delta\overline{w}_{l'}} \right \}},{\left\{ {\delta I_{s'}} \right \}}} \right )}\nonumber\\
&=&\sum_{q'}{\frac{\partial {S_l}}{\partial{\left ( {{\delta\overline{t}_{q'}}} \right )}}}\delta\overline{t}_{q'}+{\frac{\partial {S_l}}{\partial{\left ( {{\delta\overline{v}}} \right )}}}\delta\overline{v}+\sum_{l'}{\frac{\partial {S_l}}{\partial{\left ( {{\delta\overline{w}_{l'}}} \right )}}}\delta\overline{w}_{l'}+\sum_{s'}{\frac{\partial {S_l}}{\partial{\left ( {{\delta{I}_{s'}}} \right )}}}\delta{I}_{s'},\nonumber\\
&&\qquad l=1,\ldots,n.{\label{sfc}}
\end{eqnarray}
Note that every coefficient in {{{Eqs.}~{(\ref{rfc})}}} and {{(\ref{sfc})}} is a constant, which is independent of ${\left ( {{\left\{ {\delta\overline{t}_{q'}} \right \}},\delta\overline{v},{\left\{ {\delta\overline{w}_{l'}} \right \}},{\left\{ {\delta I_{s'}} \right \}}} \right )}$.

Meanwhile, from {{{Eq.}~{(\ref{i10delta})}}}, we obtain
\begin{eqnarray}
&& \delta I_{1,0}{{\left [ {t_{{0}}^\ast({k+1})} \right ]}}\nonumber\\
&=&-\frac{A_{{{\rm PP}}}}{Q}\sum_{q'}\tilde{J}_{0q'}\frac{-{e}^{-{\left [ {t_0^\ast(k+1)-t_{q'}^\ast(k)} \right ]}/\tau_{{{\rm PP}},1}}}{\tau_{{{\rm PP}},1}{\left ( {\tau_{{{\rm PP}},1}-\tau_{{{\rm PP}},2}} \right )}}\delta \overline{t}_{q'}(k)+\delta I_{1,0}{{\left [ {t_{{0}}^\ast({k})} \right ]}}{e}^{-\tilde{T}/\tau_{{{\rm IP}},1}}.\nonumber\\
&&{\label{i10deltaone}}
\end{eqnarray}
It means that $\delta I_{1,0}{{\left [ {t_{{0}}^\ast({k+1})} \right ]}}$ is a function of ${\left\{ {\delta\overline{t}_{q'}(k)} \right \}} {\mbox{ and }} \delta I_{1,0}{{\left [ {t_{{0}}^\ast({k})} \right ]}}.$
We obtain the similar relation for the rest of ${\left\{ {\delta I_{s',0}{{\left [ {t_{{0}}^\ast({k+1})} \right ]}}} \right \}}$, and they are also functions of ${\left\{ {\delta\overline{t}_{q'}(k)} \right \}}$ and $\delta I_{s',0}{{\left [ {t_{{0}}^\ast({k})} \right ]}}.$

Now, we define the vector 
\begin{eqnarray}
 \delta\tilde{\bf x}_q(k)&=&{\left ( {\delta \overline{v}_q{{\left [ {t_{{q}}^\ast({k})} \right ]}},\delta \overline{w}_{q1}{{\left [ {t_{{q}}^\ast({k})} \right ]}},\ldots,\delta \overline{w}_{qn}{{\left [ {t_{{q}}^\ast({k})} \right ]}},\delta I_{1,q}{{\left [ {t_{{q}}^\ast({k})} \right ]}},\ldots,\delta I_{4,q}{{\left [ {t_{{q}}^\ast({k})} \right ]}}} \right )},\nonumber\\
&&\qquad k={\ldots,-2,-1,0,1,2\ldots},\ q=0,\ldots,Q-1.
\end{eqnarray}
Then, noting {{{Eq.}~{(\ref{rerespiketiming})}}}, we can summarize {{{Eqs.}~{(\ref{defrf})}}}-{{(\ref{i10deltaone})}}, and so on in the form
\begin{equation}
 \delta \tilde{\bf x}_0(k+1)=\sum_{q=1}^{Q-1}{\bf A}_q\ \delta\tilde{\bf x}_q(k)+{\bf B}\ \delta\tilde{\bf x}_0(k),{\label{defab}}
\end{equation}
where the definitions of the matrices ${\bf A}_q$ and ${\bf B}$ are given in Appendix~\ref{matrixab}.
Furthermore, we define the vectors
\begin{eqnarray}
 \delta{\bf X}(0)&=&{\left ( {\delta\tilde{\bf x}_{Q-1}(k),\delta\tilde{\bf x}_{Q-2}(k),\ldots,\delta\tilde{\bf x}_0(k)} \right )},\\
 \delta{\bf X}(1)&=&{\left ( {\delta\tilde{\bf x}_{0}(k+1),\delta\tilde{\bf x}_{Q-1}(k),\ldots,\delta\tilde{\bf x}_1(k)} \right )}.
\end{eqnarray}
Then, the relation between $\delta{\bf X}(0)$ and $\delta{\bf X}(1)$ is written as
\begin{equation}
 \delta{\bf X}(1)={\bf M}\ \delta{\bf X}(0),
\end{equation}
where
\begin{equation}
{\bf M}=\left(\begin{array}{lllll}
{\bf A}_{Q-1}&{\bf A}_{Q-2}&\ldots&{\bf A}_{1}&{\bf B}\\
{\bf E}&{{\bf 0}}&\ldots&{{\bf 0}}&{{\bf 0}}\\
{{\bf 0}}&{\bf E}&&{{\bf 0}}&{{\bf 0}}\\
\vdots&&\ddots&&\vdots\\
{{\bf 0}}&{{\bf 0}}&&{\bf E}&{{\bf 0}}
\end{array} \right ).{\label{defm}}
\end{equation}
Because of the symmetrical properties of the present system, we have
\begin{equation}
 \delta{\bf X}(2)={\bf M}\ \delta{\bf X}(1)={\bf M}^2\ \delta{\bf X}(0),
\end{equation}
where
\begin{equation}
 \delta{\bf X}(2)={\left ( {\delta\tilde{\bf x}_{1}(k+1),\delta\tilde{\bf x}_{0}(k+1),\delta\tilde{\bf x}_{Q-1}(k),\ldots,\delta\tilde{\bf x}_2(k)} \right )}.
\end{equation}
Following the same scheme, we obtain the vectors of the form
\begin{equation}
  \delta{\bf X}(n)={\bf M}^n\ \delta{\bf X}(0),\qquad 1\le n,
\end{equation}
where $\delta{\bf X}(n)$ represents the deviations in the future. 

Now the stability problem of the periodic solution is reduced into the eigenvalue problem with the finite size of the matrix ${\bf M}.$
As will be shown in Appendices~\ref{matrixab} and \ref{evaluationrs}, we can easily evaluate the matrix ${\bf M}$ numerically since the coefficients in {{{Eqs.}~{(\ref{rfc})}}} and {{(\ref{sfc})}} are obtained by numerical integration of the single-body dynamics~{{(\ref{vdynev})}}-{{(\ref{iipev})}}.
In general, the matrix derived in Floquet theory always has the eigenvalue $\lambda_1=1$ with the eigenvector corresponding to the time shift in the periodic solution.
The matrix ${\bf M}$ also has the eigenvalue $\lambda_1=1$ with the eigenvector 
\begin{equation}
 \delta{\bf X}_0={\left ( {\delta\tilde{\bf x}_0,\ldots,\delta\tilde{\bf x}_0} \right )}{\label{solo}}
\end{equation}
with
\begin{equation}
 \delta\tilde{\bf x}_0={\left ( {\dot{v}_0^\ast{{\left [ {t_{{0}}^\ast({k})} \right ]}},\dot{w}_{01}^\ast{{\left [ {t_{{0}}^\ast({k})} \right ]}},\ldots,\dot{w}_{0n}^\ast{{\left [ {t_{{0}}^\ast({k})} \right ]}},\dot{I}_{1,0}^\ast{{\left [ {t_{{0}}^\ast({k})} \right ]}},\ldots,\dot{I}_{4,0}^\ast{{\left [ {t_{{0}}^\ast({k})} \right ]}}} \right )},
\end{equation}
where we define ${\left\{ {\dot{I}_{s,0}^\ast{{\left [ {t_{{0}}^\ast({k})} \right ]}}} \right \}}$ by substituting $\delta\overline{t}_{q'}(k')=0\ (k'={\ldots,-2,-1,0,1,2\ldots},\ q'=0,\ldots,Q-1)$ into the derivatives of {{{Eqs.}~{(\ref{i1})}}}-{{(\ref{i4})}}.
If the periodic solution is stable, the absolute value of other eigenvalues $\left|\lambda_m\right|\ (1<m)$ must be less than 1.
Therefore, we can determine the stability of the perfect retrieval state by numerical computation of the eigenvalues of ${\bf M}.$

In the following sections, we will apply the present analysis to evaluate the stable perfect retrieval state for the various value of parameters.
As will be shown, the present analysis is powerful enough to draw the phase diagrams.

\section{Retrieval process}{\label{sect:retrievalprocess}}

In this section, we illustrate the typical behavior of network in the process of memory retrieval.
In what follows, we always assume $Q=10$, which brings about discrete type of firing pattern of memory retrieval.
For the initial condition of the network, we set all state of neuron ${\left ( {v_i,{\left\{ {w_{il}} \right \}}} \right )}$ to be at the stable fixed point of the dynamics~{{(\ref{vdyn})}} and {{(\ref{wdyn})}} with $I_i=0$.
To evoke the retrieval of pattern~1, we give the external stimuli of the form
\begin{equation}
 I_{{{\rm EXT}},i}=\left \{
\begin{array}{lc}
A_{{{\rm EXT}}}\delta{\left ( {t-\frac{T_{{{\rm EXT}}}}{Q}q_i^1} \right )}&{\mbox{for}\ \ } 0\le t\le a_{{{\rm EXT}}}T_{{{\rm EXT}}},\\
0&\mbox{otherwise},
\end{array}\right .{\label{iext}}
\end{equation}
where $\delta (t)$ represents the delta function, and the parameters $A_{{{\rm EXT}}}$, $T_{{{\rm EXT}}}$, and $a_{{{\rm EXT}}}$ are chosen so that the initial part of the pattern~1 is forced to be retrieved.
In the present study, we set $A_{{{\rm EXT}}}=30$, $T_{{{\rm EXT}}}\sim \tilde{T}$, and $a_{{{\rm EXT}}}<0.1$.
Note that the external stimuli $I_{{{\rm EXT}},i}$ is applied only in the beginning of the network dynamics.

\begin{figure}
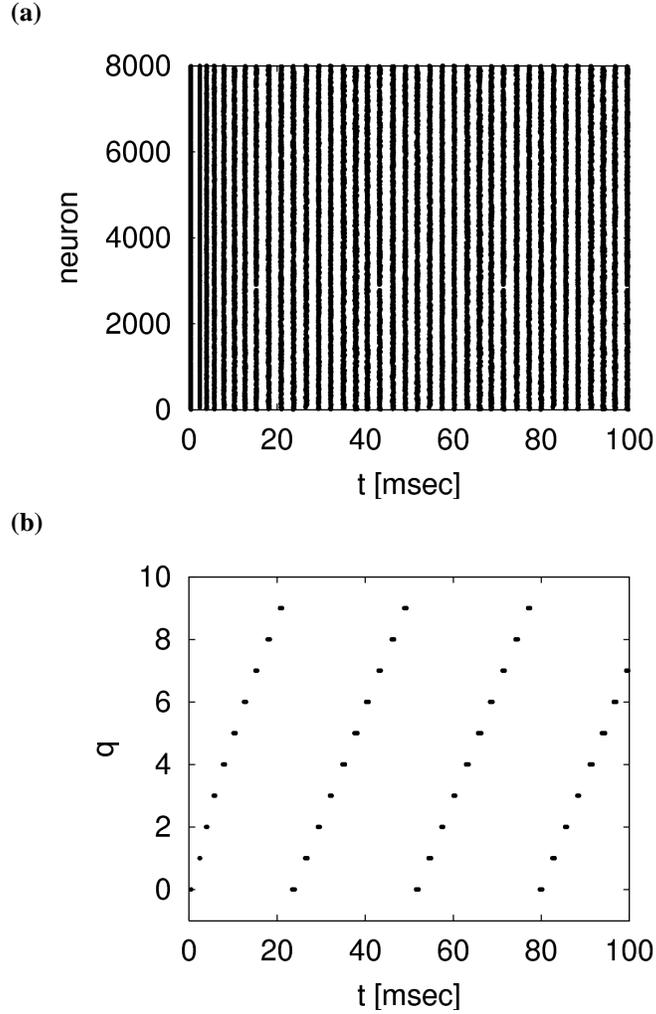

{\begin{center}\begin{tabular}{l}
{{\bf (a)}}\\
{\includegraphics[scale={0.7}]{{{d3f1}}}}\\
{{\bf (b)}}\\
{\includegraphics[scale={0.7}]{{{d3f2}}}}\\
\end{tabular}
\end{center}}
\caption{The retrieval process of pattern~1 observed in the numerical simulation with $A_{{{\rm PP}}}=30000$, $\tau_{{{\rm PP}},1}=3{\mbox{ [msec]}}$, $\tau_{{{\rm PP}},2}=0.3{\mbox{ [msec]}}$, $A_{{{\rm IP}}}=250$, $\tau_{{{\rm PP}},1}=10{\mbox{ [msec]}}$, $\tau_{{{\rm PP}},2}=1{\mbox{ [msec]}}$, $P=3$, and $N=8000$.
{{\bf (a)}}~Spike timing of neurons are plotted by closed circles as a function of time.
The initial firings of the neurons are evoked by the external electric current~$I_{{{\rm EXT}},i}$, while other firings are brought about by the synaptic electric current $I_{{{\rm PP}},i}+I_{{{\rm IP}}}$.
{{\bf (b)}}~Setting the vertical axis representing $q_i^1$ we replot the same result of the numerical simulation.
The neurons belonging to the same sublattice exhibit synchronized firing, which are expressed as overlapped closed circles.}{\label{d3f1}}
\end{figure}

In {{{Fig.}~\ref{d3f1}}}{{\bf (a)}}, we describe the result of the numerical simulation with $P=3$ and $N=8000$.
The initial firings of the neurons are evoked by the external electric current~$I_{{{\rm EXT}},i}$, while other firings are brought about by the synaptic electric current $I_{{{\rm PP}},i}+I_{{{\rm IP}}}$.
The firing pattern in {{{Fig.}~\ref{d3f1}}}{{\bf (a)}} , which looks like vertical bars, indicates the synchronized firing of a numerous number of neurons.
Since it is difficult to see whether the retrieval of pattern~1 is realized in {{{Fig.}~\ref{d3f1}}}{{\bf (a)}}, we replot the same result of numerical simulation in {{{Fig.}~\ref{d3f1}}}{{\bf (b)}}, where the vertical axes is set to represent $q_i^1$.
In this figure, we clearly see the successful retrieval of pattern~1, in which the neurons belonging to the same sublattice exhibit synchronized firing.

\begin{figure}
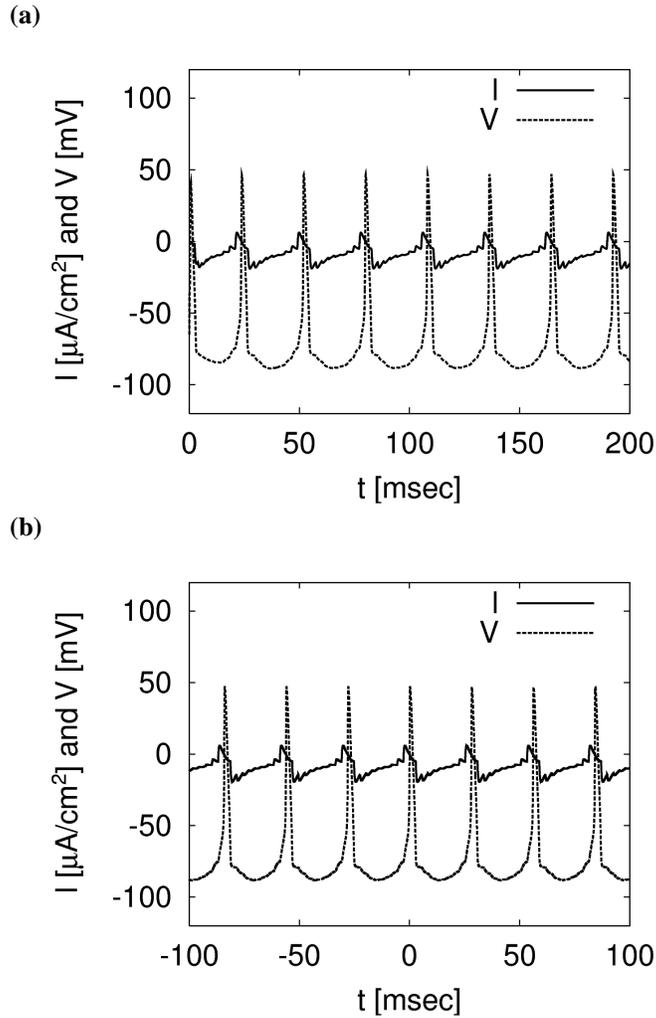

{\begin{center}\begin{tabular}{l}
{{\bf (a)}}\\
{\includegraphics[scale={0.7}]{{{d3}}}}\\
{{\bf (b)}}\\
{\includegraphics[scale={0.7}]{{{d3t}}}}\\
\end{tabular}
\end{center}}
\caption{{{\bf (a)}}\ The behavior of a neuron with $q_i^1=0$ observed in the numerical simulation in {{{Fig.}~\ref{d3f1}}}.
The membrane potential $v_i$ and the synaptic electric current $I_i$ are plotted as a function of time.
{{\bf (b)}}\ The result of the theoretical analysis for the stationary state of the neuron, which explains the stationary behavior of the neuron in {{\bf (a)}}.}{\label{d3}}
\end{figure}

The dynamical behavior of the neuron with $q_i^1=0$ is described in {{{Fig.}~\ref{d3}}}{{\bf (a)}} .
After the transient behavior, the neuron settles into the stationary state, where the neuron exhibits periodic firing.
In {{{Fig.}~\ref{d3}}}{{\bf (b)}}, we describe the periodic solution for retrieval state obtained from {{{Eq.}~{(\ref{conditionr})}}}.
In order to examine the stability of this solution, we calculate the explicit value of the matrix ${\bf M}$ numerically.
In the present case, the largest absolute eigenvalue is 1, and the theoretically evaluated perfect retrieval state in {{{Fig.}~\ref{d3}}}{{\bf (b)}}\ is stable.
The good agreement between {{{Fig.}~\ref{d3}}}{{\bf (a)}}\ and {{\bf (b)}}\ implies the validity of the present analysis.
It is also worth noting that the theoretical result in {{{Fig.}~\ref{d3}}}{{\bf (b)}}\ is independent of $P$ because of {{{Eq.}~{(\ref{vanish})}}}.
We set $P=3$ in the numerical simulation in {{{Fig.}~\ref{d3}}}{{\bf (a)}}, and this result of numerical simulation is well explained by the $P$-independent solution in {{{Fig.}~\ref{d3}}}{{\bf (b)}}.
We will see the same result of numerical simulation even with the larger value of $P$, as far as $P/N$ is sufficiently small.

\section{Phase transition due to the loss of the stability of the perfect retrieval state}{\label{sect:lossofstability}}

\begin{figure}
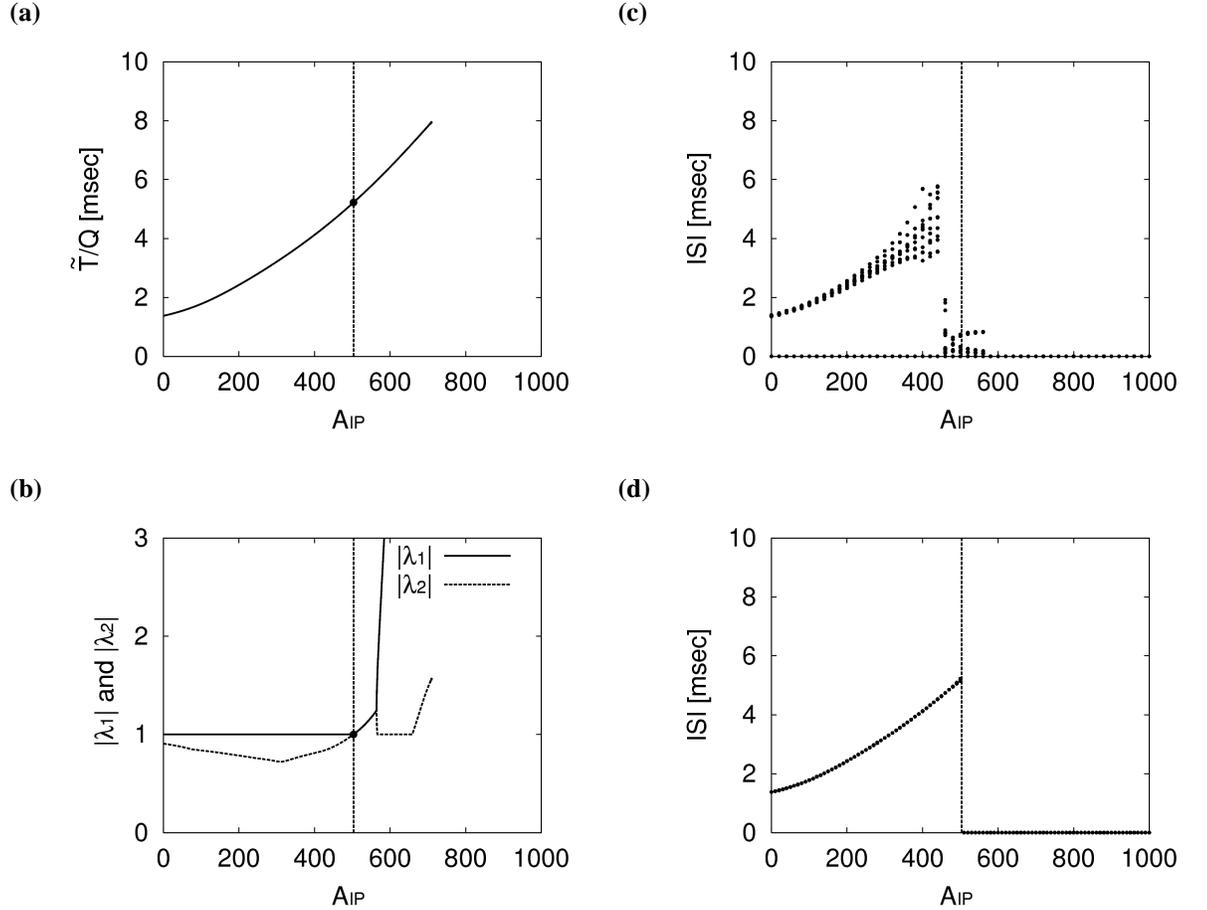

{\begin{center}\begin{tabular}{ll}
{{\bf (a)}} & {{\bf (c)}}\\
{\includegraphics[scale={0.6}]{{{d3l}}}}&{\includegraphics[scale={0.6}]{{{d3n8}}}}\\
&\\
{{\bf (b)}} & {{\bf (d)}}\\
{\includegraphics[scale={0.6}]{{{d3la}}}}&{\includegraphics[scale={0.6}]{{{d3sl}}}}
\end{tabular}\end{center}}
\caption{
{{\bf (a)}}~$\tilde{T}/Q$ obtained from {{{Eq.}~{(\ref{conditionr})}}} is plotted as a function of $A_{{{\rm IP}}}$.
{{\bf (b)}}~The largest two absolute eigenvalues of the matrix ${\bf M}$ are plotted as a function of $A_{{{\rm IP}}}$.
One of the absolute eigenvalues exceeds 1 beyond the critical point~$A_{{{\rm PP}}}^c\sim 500$, which is represented by the vertical lines in all four figures.
The perfect retrieval state we have evaluated in {{\bf (a)}}\ is stable only below the critical point~$A_{{{\rm PP}}}^c$.
{{\bf (c)}}~Changing the value of $A_{{{\rm IP}}}$ we observe the ISIs of all neurons in the numerical simulations with $P=1$ and $N=8000$.
See text for the definition of the ISIs we calculate here.
{{\bf (d)}}~We observe the ISIs also in the dynamics of sublattices~{{(\ref{vdynsl})}}-{{(\ref{isumsl})}}.
The critical point~$A_{{{\rm IP}}}^c$ evaluated in {{\bf (a)}}\ and {{\bf (b)}}\ explains the phase transition observed in {{\bf (c)}}\ and {{\bf (d)}}\ well, although we see some fluctuations near $A_{{{\rm IP}}}^c$ owing to the finite number of neurons in the case of {{\bf (c)}}.
Beyond the critical point~$A_{{{\rm IP}}}^c$ the network settles into another stationary state.
The value of parameters are $\tau_{{{\rm PP}},1}=3{\mbox{ [msec]}}$, $\tau_{{{\rm PP}},2}=0.3{\mbox{ [msec]}}$, $A_{{{\rm PP}}}=30000$, $\tau_{{{\rm IP}},1}=10{\mbox{ [msec]}}$, and $\tau_{{{\rm IP}},2}=1{\mbox{ [msec]}}$.
}{\label{d3l}}
\end{figure}

Here we investigate the effect of inhibitory synaptic electric current $I_{{{\rm IP}}}$, which is controlled by $A_{{{\rm IP}}}$.
From {{{Eq.}~{(\ref{conditionr})}}}, we obtain $\tilde{T}/Q$ as a function of $A_{{{\rm IP}}}$, which is plotted in {{{Fig.}~\ref{d3l}}}{{\bf (a)}} .
As $A_{{{\rm IP}}}$ increases, the period of the retrieval process $\tilde{T}$ becomes longer since each neuron obtains a large amount of inhibitory synaptic electric $I_{{{\rm IP}}}$ with the large value of $A_{{{\rm IP}}}$.
Figure~\ref{d3l}{{\bf (b)}}\ describes the absolute eigenvalues of the matrix ${\bf M}$.
Size of the matrix~${\bf M}$ is $80\times 80$, and the largest two absolute eigenvalues are plotted in {{{Fig.}~\ref{d3l}}}{{\bf (b)}} .
With $A_{{{\rm IP}}}{{\raisebox{-1ex}{\mbox{$\stackrel{{<}}{\sim}$}}}} 500$ the largest absolute eigenvalue is 1, while it exceeds 1 with $500{{\raisebox{-1ex}{\mbox{$\stackrel{{<}}{\sim}$}}}} A_{{{\rm IP}}}$, that is to say, the stability of the perfect retrieval state is lost beyond the critical point $A_{{{\rm IP}}}^c\sim 500$.

To observe this phase transition in numerical simulations, we calculate the inter spike intervals (ISIs) of all neurons changing the value of $A_{{{\rm IP}}}$, as described in  {{{Fig.}~\ref{d3l}}}{{\bf (c)}}.
Note that the ISIs we calculate here is based on spike timing of all neurons.
When neuron~$i$ and neuron~$j$ fire sequentially at time $t_i$ and $t_j$ respectively, we calculate the time difference $t_j-t_i$ to obtain the ISIs of all neurons.
By means of these ISIs, we can evaluate the gaps in spike timing appearing in {{{Fig.}~\ref{d3f1}}}{{\bf (a)}}, which corresponds to $\tilde{T}/Q$.
The theoretical result in {{{Fig.}~\ref{d3l}}}{{\bf (a)}}\ and {{\bf (b)}}\ explains ISIs in {{{Fig.}~\ref{d3l}}}{{\bf (c)}}\ well, although we see some fluctuations due to the finite number of neurons near the critical point~$A_{{{\rm IP}}}^c$.
In {{{Fig.}~\ref{d3l}}}{{\bf (d)}}, we calculate the ISIs in the dynamics of sublattices~{{(\ref{vdynsl})}}-{{(\ref{isumsl})}}, in which we have taken the limit of the infinite number of neurons.
The theoretically evaluated critical point~$A_{{{\rm IP}}}^c$ explains the loss of the stability observed in {{{Fig.}~\ref{d3l}}}{{\bf (d)}}\ with a high degree of precision.

\begin{figure}
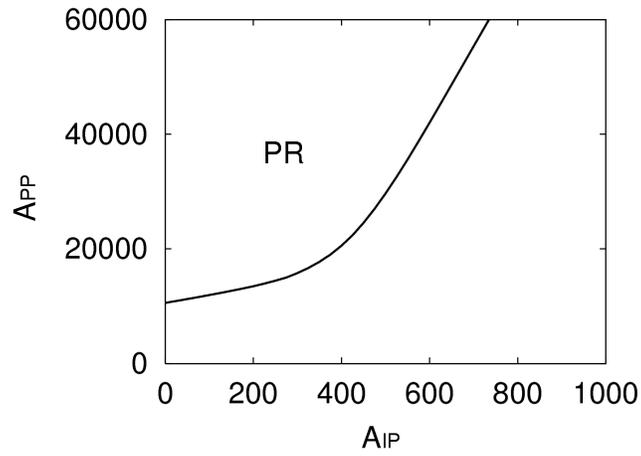

{\begin{center}
{\includegraphics[scale={0.7}]{{{pd3}}}}
\end{center}}
\caption{$A_{{{\rm IP}}}-A_{{{\rm PP}}}$ phase diagram obtained by the theoretical analysis.
The stable perfect retrieval state is found in the region denoted by PR.
The value of parameters are $\tau_{{{\rm PP}},1}=3{\mbox{ [msec]}}$, $\tau_{{{\rm PP}},2}=0.3{\mbox{ [msec]}}$, $\tau_{{{\rm IP}},1}=10{\mbox{ [msec]}}$, and $\tau_{{{\rm PP}},1}=1{\mbox{ [msec]}}$.}{\label{pd3}}
\end{figure}

In {{{Fig.}~\ref{pd3}}}, we draw $A_{{{\rm IP}}}$-$A_{{{\rm PP}}}$ phase diagram, which is evaluated by the theoretical analysis.
We find the stable perfect retrieval state in the region represented by PR.
As $A_{{{\rm PP}}}$ decreases, the range of $A_{{{\rm IP}}}$ for the stable perfect retrieval state becomes narrower since a large amount of $I_{{{\rm PP}}}$ is required for the successful memory retrieval under the strong inhibition.

\section{Two separated perfect retrieval phases appearing with the slow $\alpha$-function $S_{{{\rm PP}}}(t)$}{\label{sect:separated}}

\begin{figure}
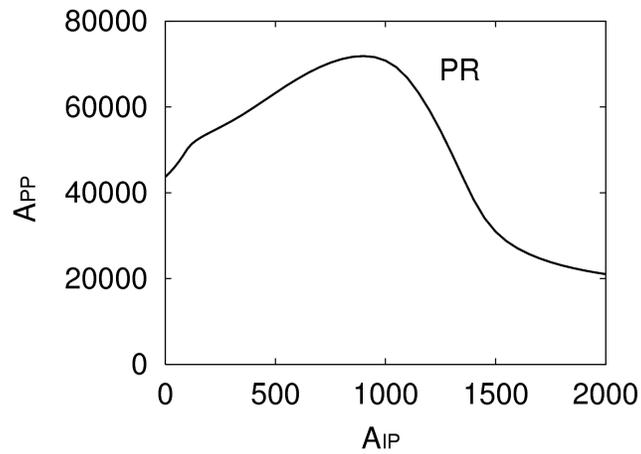

{\begin{center}
{\includegraphics[scale={0.7}]{{{pd20}}}}
\end{center}}
\caption{The same as {{{Fig.}~\ref{pd3}}}, except that $\tau_{{{\rm PP}},1}=20{\mbox{ [msec]}}$ and $\tau_{{{\rm PP}},2}=2{\mbox{ [msec]}}$.}{\label{pd20}}
\end{figure}

In the previous sections, we assume $S_{{{\rm PP}}}(t)$ with $\tau_{{{\rm PP}},1}=3{\mbox{ [msec]}}$ and $\tau_{{{\rm PP}},2}=0.3{\mbox{ [msec]}}$ as well as $S_{{{\rm IP}}}(t)$ with $\tau_{{{\rm IP}},1}=10{\mbox{ [msec]}}$ and $\tau_{{{\rm IP}},2}=1{\mbox{ [msec]}}$, where $S_{{{\rm PP}}}(t)$ decays much faster than $S_{{{\rm IP}}}(t)$.
In order to examine the role of the decay time constants in $\alpha$-functions, we investigate the case of the slow $\alpha$-function $S_{{{\rm PP}}}(t)$ with $\tau_{{{\rm PP}},1}=20{\mbox{ [msec]}}$ and $S_{{{\rm PP}},2}=2{\mbox{ [msec]}}$.
For this slow $\alpha$-function~$S_{{{\rm PP}}}(t)$, we describe $A_{{{\rm IP}}}-A_{{{\rm PP}}}$~phase diagram in {{{Fig.}~\ref{pd20}}}.
The distinctive feature of this phase diagram is the perfect retrieval phase appearing in the region with the large value of $A_{{{\rm IP}}}$.
In the case of the fast $\alpha$-function~$S_{{{\rm PP}}}(t)$, the strong inhibition with the large value of $A_{{{\rm IP}}}$ tends to suppress the perfect retrieval, as described in {{{Fig.}~\ref{pd3}}}.
Nevertheless, in {{{Fig.}~\ref{pd20}}}, we see two separated perfect retrieval phases in the region with $40000{{\raisebox{-1ex}{\mbox{$\stackrel{{<}}{\sim}$}}}} A_{{{\rm PP}}}{{\raisebox{-1ex}{\mbox{$\stackrel{{<}}{\sim}$}}}} 70000$, while these two retrieval phases merge with each other in the region with $70000{{\raisebox{-1ex}{\mbox{$\stackrel{{<}}{\sim}$}}}} A_{{{\rm PP}}}$.

\begin{figure}
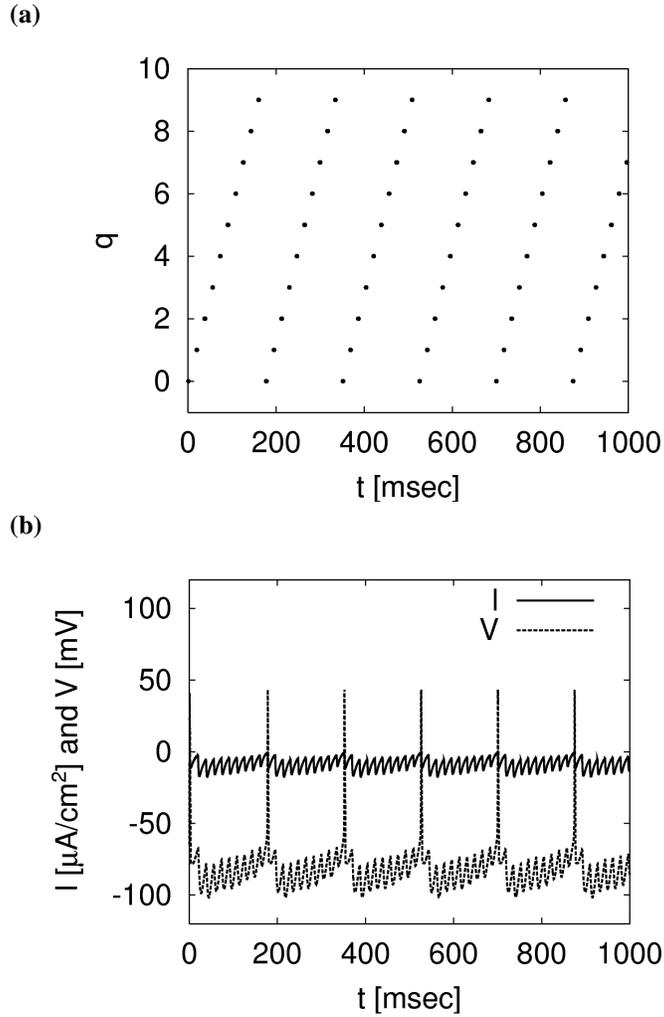

{\begin{center}\begin{tabular}{l}
{{\bf (a)}}\\
{\includegraphics[scale={0.7}]{{{d20f2}}}}\\
{{\bf (b)}}\\
{\includegraphics[scale={0.7}]{{{d20}}}}\\
\end{tabular}
\end{center}}
\caption{The retrieval process of pattern~1 in the case of the slow $\alpha$-function~$S_{{{\rm PP}}}(t)$ with $\tau_{{{\rm PP}},1}=20{\mbox{ [msec]}}$ and $\tau_{{{\rm PP}},2}=2{\mbox{ [msec]}}$ under the strong inhibition with $A_{{{\rm IP}}}=1500$.
{{\bf (a)}}~Spike timing of neurons observed in the numerical simulations with $P=3$ and $N=8000$ are plotted as a function of time.
Note that the vertical axis represents $q_i^1$.
{{\bf (b)}}~The membrane potential~$v_i$ and the synaptic electric current~$I_i$ are plotted as a function of time for a neuron with $q_i^1=0$.
The value of parameters are  $A_{{{\rm PP}}}=65000$, $\tau_{{{\rm IP}},1}=10{\mbox{ [msec]}}$, and $\tau_{{{\rm IP}},2}=1{\mbox{ [msec]}}$.}{\label{d20f2}}
\end{figure}

One example of the retrieval process in the region with the large value of $A_{{{\rm IP}}}$ is illustrated in {{{Fig.}~\ref{d20f2}}}.
As a result of the large value of $A_{{{\rm IP}}}$, the neuron obtains a large amount of the inhibitory electric current $I_{{{\rm IP}}}$, which oscillates with the period $\tilde{T}/Q$.
In the retrieval process, $Q$ sublattices emerge exhibiting synchronized firing of neurons, as described in {{{Fig.}~\ref{d20f2}}}{{\bf (a)}}.
When one firing of sublattice occurs, all neurons obtain a large amount of the inhibitory synaptic electric current~$I_{{{\rm IP}}}$.
Then, neurons in the next sublattice cannot fire until this inhibitory electric current decays with the time constant~$\tau_{{{\rm IP}},1}$.
In this way, the oscillatory inhibitory electric current~$I_{{{\rm IP}}}$ regulates the spike intervals of sublattices, and hence the memory retrieval with the long period~$\tilde{T}$ is realized.
The long-time influence of the slow $\alpha$-function $S_{{{\rm PP}}}(t)$ is indispensable for this memory retrieval since the time gaps of firings of sublattices ({i.e.}, $\tilde{T}/Q$) are considerably large.

\begin{figure}
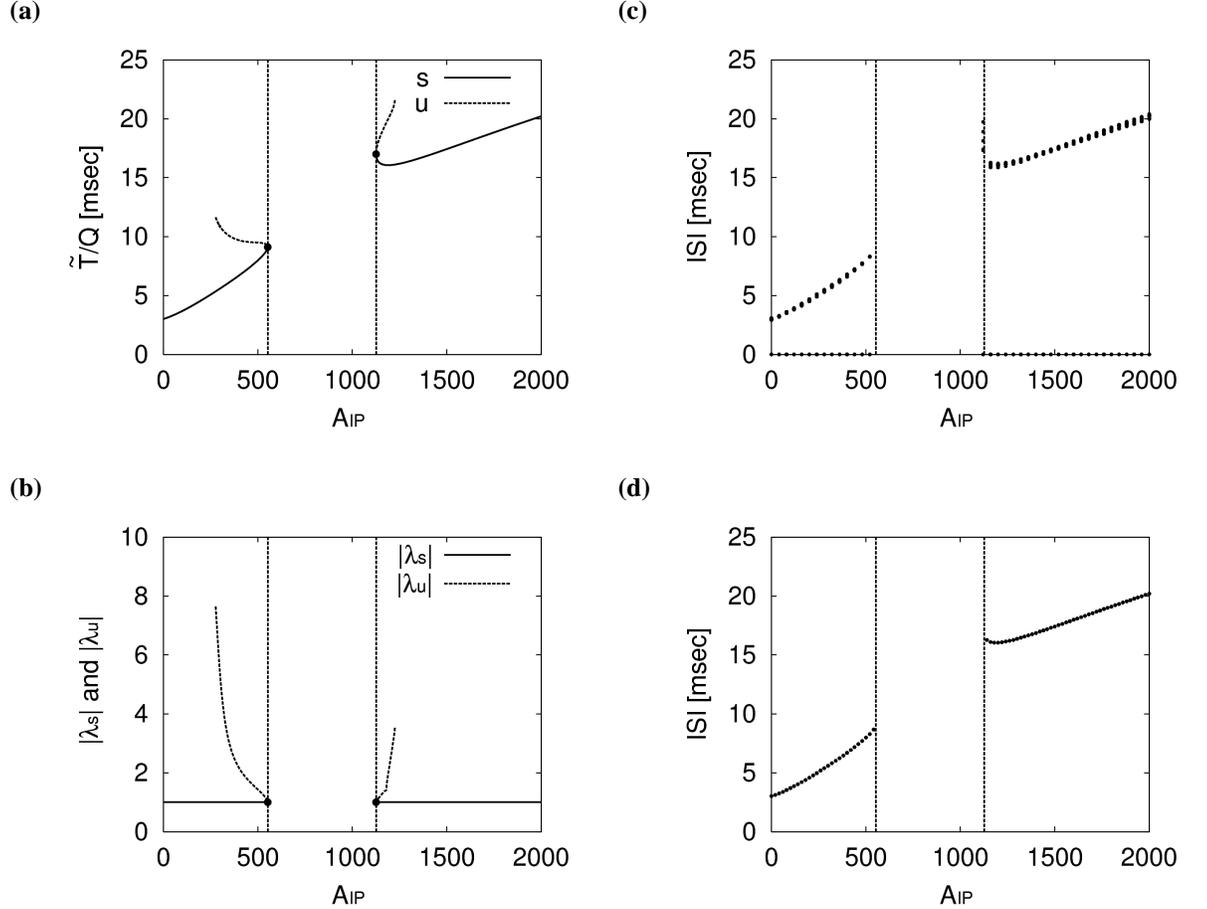

{\begin{center}\begin{tabular}{ll}
{{\bf (a)}} & {{\bf (c)}}\\
{\includegraphics[scale={0.6}]{{{d20l}}}}&{\includegraphics[scale={0.6}]{{{d20n8}}}}\\
&\\
{{\bf (b)}} & {{\bf (d)}}\\
{\includegraphics[scale={0.6}]{{{d20la}}}}&{\includegraphics[scale={0.6}]{{{d20sl}}}}
\end{tabular}\end{center}}
\caption{{{\bf (a)}}~For the case of the slow $\alpha$-function $S_{{{\rm PP}}}(t)$ with $\tau_{{{\rm PP}},1}=20{\mbox{ [msec]}}$ and $\tau_{{{\rm PP}},2}=2{\mbox{ [msec]}}$, we plot $\tilde{T}/Q$ obtained from {{{Eq.}~{(\ref{conditionr})}}} as a function of $A_{{{\rm IP}}}$.
Near the boundary, we find two solutions of {{{Eq.}~{(\ref{conditionr})}}}, which are represented by `s' and `u'.
{{\bf (b)}}~The largest absolute eigenvalue of the matrix~${\bf M}$ is plotted as a function~$A_{{{\rm IP}}}$.
The eigenvalues for state~s and state~u are represented by $\lambda_s$ and $\lambda_u$, respectively.
This result implies that state~s is stable, while state~u is unstable.
The two critical point~$A_{{{\rm IP}}}^c(1)$ and $A_{{{\rm IP}}}^c(2)$ obtained from the theoretical analysis {{\bf (a)}}\ and {{\bf (b)}}\ are represented by the vertical lines in all four figures.
{{\bf (c)}}~The ISIs observed in the numerical simulations with $P=1$ and $N=8000$.
We slowly change the value of $A_{{{\rm IP}}}$ both from $A_{{{\rm IP}}}=0$ and $A_{{{\rm IP}}}=2000$.
{{\bf (d)}}~The ISIs observed in the dynamics of sublattices~{{(\ref{vdynsl})}}-{{(\ref{isumsl})}}.
The phase transitions observed in the numerical simulations {{\bf (c)}}\ and {{\bf (d)}}\ are well explained by the critical points~$A_{{{\rm IP}}}^c(1)$ and $A_{{{\rm IP}}}^c(2)$.
In {{\bf (c)}}\ and {{\bf (d)}}, all neurons cease firing between $A_{{{\rm IP}}}^c(1)$ and $A_{{{\rm IP}}}^c(2)$.
The value of parameters are $A_{{{\rm PP}}}=65000$, $\tau_{{{\rm IP}},1}=10{\mbox{ [msec]}}$, and $\tau_{{{\rm IP}},2}=1{\mbox{ [msec]}}$.
}{\label{d20l}}
\end{figure}

In {{{Fig.}~\ref{d20l}}} {{\bf (a)}}\ and {{\bf (b)}}, we describe the result of our analysis with $A_{{{\rm PP}}}=65000$, where we see two separated perfect retrieval phases.
Near the boundary of the retrieval phases, {{{Eq.}~{(\ref{conditionr})}}} yields two different perfect retrieval states, which are indicated by `s' and `u' in {{{Fig.}~\ref{d20l}}} {{\bf (a)}}.
As described in {{{Fig.}~\ref{d20l}}} {{\bf (b)}}, the largest absolute eigenvalue of the matrix ${\bf M}$ for the state~u exceeds 1, while that for the state~s always takes 1.
This result implies that the state~s is stable and the state~u is unstable.
The ISIs observed in the numerical simulations are plotted as a function of $A_{{{\rm IP}}}$ in {{{Fig.}~\ref{d20l}}} {{\bf (c)}}.
To obtain these ISIs, we slowly change the value of $A_{{{\rm IP}}}$ both from $A_{{{\rm IP}}}=0$ and from $A_{{{\rm IP}}}=2000$.
In the present case, neurons cease firing between the critical points~$A_{{{\rm IP}}}^c(1)$ and $A_{{{\rm IP}}}^c(2)$.
The ISIs observed in the dynamics of sublattices~{{(\ref{vdynsl})}}-{{(\ref{isumsl})}} are also plotted in {{{Fig.}~\ref{d20l}}} {{\bf (d)}}.
The phase transitions observed in {{{Fig.}~\ref{d20l}}} {{\bf (c)}}\ and {{\bf (d)}}\ are well explained by the theoretical analysis in {{{Fig.}~\ref{d20l}}} {{\bf (a)}}\ and {{\bf (b)}}.

\begin{figure}
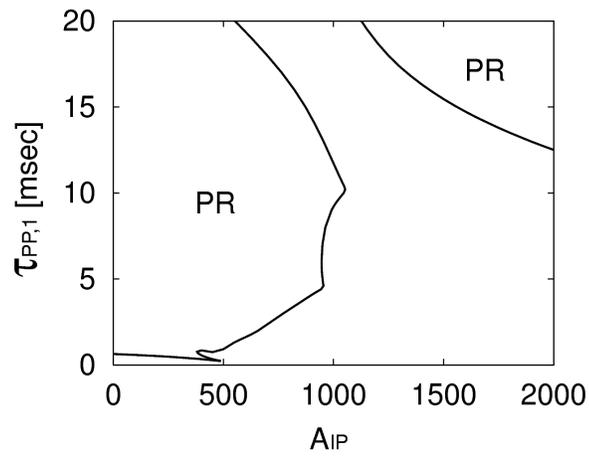

{\begin{center}
{\includegraphics[scale={0.7}]{{{pdx}}}}
\end{center}}
\caption{$A_{{{\rm IP}}}-\tau_{{{\rm PP}},1}$ phase diagram obtained by the theoretical analysis, where we fix $\tau_{{{\rm PP}},2}=0.1\tau_{{{\rm PP}},1}$.
The value of parameters are $A_{{{\rm PP}}}=65000$, $\tau_{{{\rm IP}},1}=10{\mbox{ [msec]}}$, and $\tau_{{{\rm IP}},2}=1{\mbox{ [msec]}}$.}{\label{pdx}}
\end{figure}

In order to investigate more details about decay time constants, we describe $A_{{{\rm IP}}}-\tau_{{{\rm PP}},1}$ phase diagram in {{{Fig.}~\ref{pdx}}}, in which we fix $\tau_{{{\rm PP}},2}=0.1\tau_{{{\rm PP}},1}$.
In the region with the long $\tau_{{{\rm PP}},1}$, we find two separated retrieval phases.
In this case, stable and unstable solutions of {{{Eq.}~{(\ref{conditionr})}}} are found only inside the perfect retrieval phase, as described in {{{Fig.}~\ref{d20l}}}.
One might thus conceive that the stability analysis is not necessary for the purpose of determining the phase boundary.
However, with the short $\tau_{{{\rm PP}},1}$, we find the unstable solution of {{{Eq.}~{(\ref{conditionr})}}} outside the perfect retrieval phase, as described in {{{Fig.}~\ref{d3l}}}.
The stability analysis is hence indispensable to determine the boundary of the perfect retrieval phase, particularly with the short $\tau_{{{\rm PP}},1}$.

\section{Discussion}{\label{sect:discussion}}

We have investigated associative memory neural networks of spiking neurons memorizing periodic spatio-temporal patterns of spike timing.
In encoding the multiple spatio-temporal patterns, we assume the spike-timing-dependent synaptic plasticity with the asymmetric time window $W{\left ( {\Delta t} \right )}$ in {{{Fig.}~\ref{figw}}}.
Encoded periodic spatio-temporal patterns of spike timing are reproduced successfully in the periodic firing pattern of neurons in the process of memory retrieval.
In this retrieval process, $Q$ sublattices (clusters of neurons) exhibit synchronized firing, and the oscillatory inhibitory electric current~$I_{{{\rm IP}}}$, which is supposed to come from interneurons, regulates the spike timing of sublattices.

In order to investigate the stationary properties of the system, we have derived the periodic solution for the retrieval state analytically in the limit of infinite number of neurons.
From this analysis, we have shown that if the average of the time window $W{\left ( {\Delta t} \right )}$ takes the value of zero, the crosstalk among encoded patterns vanishes.
This result implies that the present form of the time window $W{\left ( {\Delta t} \right )}$, which is found in experiments, has the great advantage in encoding a large number of spatio-temporal patterns.

To elucidate the stability of the derived periodic solution we have employed a linear stability analysis.
In this linear stability analysis we have to evaluate the time evolution of infinitesimal deviation so as to obtain the matrix for Floquet theory, although the naive application of Floquet theory yields infinite size of matrix.
In order to reduce the size of matrix, we have employed some decomposition of the stability problem, by which the original stability problem with $N$ neurons is reduced into the stability problem with $Q$ sublattices.
Then, to take account of the infinite long-time influence of $\alpha$-functions, we have introduced the variables ${\left\{ {I_{s',q}} \right \}}$, which enable us to obtain the finite size of matrix ${\bf M}$ for Floquet theory.
The explicit form of ${\bf M}$ is computed by the numerical integration of the single-body dynamics~{{(\ref{vdynev})}}-{{(\ref{iipev})}}, and the stability of solutions are evaluated from the eigenvalues of ${\bf M}$.

Based on these method of analysis, we have investigated the stationary properties of retrieval state in the case of the fast $\alpha$-function~$S_{{{\rm PP}}}(t)$ with $\tau_{{{\rm PP}},1}=3{\mbox{ [msec]}}$ and $\tau_{{{\rm PP}},2}=0.3{\mbox{ [msec]}}$.
In {{{Fig.}~\ref{d3l}}}{{\bf (a)}}, we have obtained periodic solutions for the retrieval state for various value of $A_{{{\rm PP}}}$ by solving {{{Eq.}~{(\ref{conditionr})}}}.
Then, in {{{Fig.}~\ref{d3l}}}{{\bf (b)}}, we have employed the stability analysis of these periodic solution to obtain the critical point~$A_{{{\rm IP}}}^c$ .
The phase transition observed in the numerical simulations in {{{Fig.}~\ref{d3l}}}{{\bf (c)}}\ and {{\bf (d)}}\ is well explained by this critical point~$A_{{{\rm IP}}}^c$.
The condition for the successful memory retrieval is summarized as $A_{{{\rm IP}}}-A_{{{\rm PP}}}$ phase diagram in {{{Fig.}~\ref{pd3}}}.

Meanwhile, with the slow $\alpha$-function~$S_{{{\rm PP}}}(t)$ with $\tau_{{{\rm PP}},1}=20{\mbox{ [msec]}}$ and $\tau_{{{\rm PP}},2}=2{\mbox{ [msec]}}$, we have found two separated retrieval phases, as shown in {{{Fig.}~\ref{pd20}}}.
The behavior of neurons in the memory retrieval with the large value of $A_{{{\rm IP}}}$ is described in {{{Fig.}~\ref{d20f2}}}, where we see the large size of oscillatory inhibitory synaptic electric current $I_{{{\rm IP}}}$ regulating the spike timing of neurons.
The result of the theoretical analysis is illustrated in {{{Eq.}~{(\ref{d3l})}}}, where the stability analysis is used to chose the stable solution from the multiple solutions of {{{Eq.}~{(\ref{conditionr})}}}.

The heart of the present stability analysis lies in the exact reduction of the size of the matrix for Floquet theory.
Since $Q$ sublattices arise in the stationary state, we have to evaluate the matrix with the dimension of $(1+n+4)Q$, where $1+n$ corresponds to the degree of freedom of the neuron dynamics $f{\left ( {v_i,w_{i1},\ldots,w_{in}} \right )}$ and $g_l{\left ( {v_i,w_{i1},\ldots,w_{in}} \right )}$ $(l=1,\ldots,n)$, and additional 4 is required to evaluate the infinite long-time influence of $\alpha$-functions $S_{{{\rm PP}}}(t)$ and $S_{{{\rm IP}}}(t)$.
In the present study we set $Q=10$ and $n=3$, which yields the matrix with the dimension of $80$.
Although one might conceive that the size of this matrix is somewhat large, the critical points obtained from this matrix well explain the result of numerical simulations with a high degree of precision, as demonstrated in Figs.~\ref{d3l} and \ref{d20l}.
In other network models\cite{gerstner4,vreeswijk}, only a few number of sublattices emerges and the size of matrix becomes small.

The spatio-temporal patterns to be memorized are assumed to be periodic in the present study for ease of analysis.
It is worth noting that learning rule based on the spike-timing-dependent synaptic plasticity is applicable to a wide class of spatio-temporal patterns of spike timing.
The periodicity of spatio-temporal patterns is not crucial, and it is almost obvious that spike trains generated by independent Poisson process are also well encoded by use of the time window~$W{\left ( {\Delta t} \right )}$.
In the case of Poisson process, the firing rate in Poisson process must be adequately low since the refractoriness of neurons is expected to prevent retrieval of Poisson trains with high firing rate. 

It is of interest to consider the effect of noise in the present model.
With the large value of $A_{{{\rm IP}}}$, neurons obtain the large size of oscillatory inhibitory electric current~$I_{{{\rm PP}}}$ as described in {{{Fig.}~\ref{d20f2}}}, and the effect like stochastic resonance is expected to occur in the presence of noise.
The evaluation of the effect of noise, however, seems to be difficult in the present scheme of analysis since we are required to calculate the distribution of spike timing of neurons in this evaluation.

With the large $A_{{{\rm IP}}}$ and the short $\tau_{{{\rm PP}},1}$ the basin of attractors for spatio-temporal patterns are found to be narrow in the present model (data not shown).
In the initial condition, the inhibitory synaptic electric current $I_{{{\rm IP}}}$ is taken to be the value of zero.
With the short $\tau_{{{\rm PP}},1}$, firing of the first sublattice, which is induced by $I_{{{\rm EXT}},i}$, brings about firing of the second sublattice immediately since some accumulation of inhibitory synaptic electric currents~$I_{{{\rm IP}}}$ are necessary to control the next firing.
For these reasons, a first few firings of sublattices take place quite rapidly.
These rapid firings of sublattices give rise to too much accumulation of inhibitory electric current~$I_{{{\rm IP}}}$, and then terminate firings of all neurons.
The core of the problem in this phenomenon is too rapid firings of interneurons.
To avoid this problem, more sophisticated modeling of interneurons is needed so as to realize adequate control of interspike intervals of interneurons.
When we assume that interneurons exhibit periodic firing independently of pyramidal neurons, the inhibitory synaptic electric current~$I_{{{\rm IP}}}$ takes the form 
\begin{equation}
 I_{{{\rm IP}}}=A_{{{\rm IP}}}\sum_k S_{{{\rm IP}}}{\left ( {t-kT_{{{\rm IP}}}} \right )},
\end{equation}
where $T_{{{\rm IP}}}$ represents the period of firings of interneurons.
We can investigate the case of this periodic inhibitory electric current $I_{{{\rm IP}}}$ following the almost same scheme of the present analysis.

Finally, we discuss the biological implication of the present study.
The result of the present study strongly suggests the possibility of the concept of temporal coding, in which information is assumed to be processed based on spike timing of neurons.
The question then arises about where we can find these kind of information processing in the real nervous system.
It is well known that the hippocampus is the important tissue for short term memory.
In CA3 region of hippocampus, we see dense recurrent connections among pyramidal neurons, and hence the short term memory is thought to be stored in the CA3 region of hippocampus.
Memory stored in hippocampus should be transfered into other regions such as neocortex so that it is stored as the long term memory.
Recently, some experimental results begin to suggest that this memory transfer process takes place when sharp waves (SPW) appear in hippocampus\cite{siapas}.
In SPW, fast periodic firings of interneurons ($\sim$200 [Hz]) bring about oscillatory inhibitory synaptic electric currents in pyramidal neurons, and these oscillatory electric currents regulate occasional firings of pyramidal neurons\cite{ylinen}.
N{\'a}dasdy {et al}. have investigated these occasional firings of pyramidal neurons and revealed that repeating firing patterns of pyramidal neurons are present in SPW\cite{nadasdy}.
These results of experiments indicate that spike timing of pyramidal neurons of SPW represent some kind of memory that should be transfered into neocortex.
Also in the gamma oscillation, we observe oscillatory inhibitory synaptic electric currents due to periodic firing of interneurons, although its frequency is somewhat low (20-80 [Hz]).
Buzs{\'a}ki {et al}. have hypothesized that the firing patterns of pyramidal neurons in the gamma oscillation are stored in the recurrent connections of the CA3 region of hippocampus, and then these stored firing patterns are replayed  in the firing patterns in SPW in the time compressed manner\cite{buzsaki2}.
Our theoretical model explains these time compressed replay of firing patterns.

Some aspects of our theoretical model are, however,  still biologically implausible.
For example, the learning rule~{{(\ref{learningrule})}} gives either negative or positive synaptic synaptic weights by chance although synaptic weight among pyramidal neurons are found to be positive in experiments.
More precise modeling of interneurons might be needed to acquire a deeper understanding of the time compressed replay of firing patterns.
Solving these problems will be part of our future study.

\appendix

\section{The Hodgkin-Huxley equations}{\label{hhequations}}

The Hodgkin-Huxley equations are the ordinary differential equations with four degrees of freedom, which have been developed to describe the spike generation of the squid's giant axon\cite{hodgkin}.
In the present study, for the dynamics $f{\left ( {v,w_{1},\ldots,w_{n}} \right )}$ and $g_l {\left ( {v,w_{1},\dots,w_{n}} \right )}\ (l=1,\ldots,n)$, we assume the Hodgkin-Huxley equations of the form
\begin{eqnarray}
C_m\ f{\left ( {v,w_1,\ldots,w_3} \right )}&=&\overline{g}_{Na}w_2^3w_1{\left ( {v_{Na}-v} \right )}+\overline{g}_{K}w_3^4{\left ( {v_{K}-v} \right )}\nonumber\\
&&+\overline{g}_{L}{\left ( {v_{L}-v} \right )},\\
g_1 {\left ( {v,w_1,\ldots,w_3} \right )}&=&\alpha_1{\left ( {1-w_1} \right )}-\beta_1 w_1,\\
g_2 {\left ( {v,w_1,\ldots,w_3} \right )}&=&\alpha_2{\left ( {1-w_2} \right )}-\beta_2 w_2,\\
g_3 {\left ( {v,w_1,\ldots,w_3} \right )}&=&\alpha_3{\left ( {1-w_3} \right )}-\beta_3 w_3
\end{eqnarray}
with
\begin{eqnarray}
\alpha_1&=&0.01{\left ( {10-v} \right )}\left /{\left\{ {\exp{\left ( {\frac{10-v}{10}} \right )}-1} \right \}}\right . ,\\
\beta_1&=&0.125\exp{\left ( {-v/80} \right )},\\
\alpha_2&=&0.1{\left ( {25-v} \right )}\left /{\left\{ {\exp{\left ( {\frac{25-v}{10}} \right )}-1} \right \}}\right . ,\\
\beta_2&=&4\exp{\left ( {-v/18} \right )},\\
\alpha_3&=&0.07\exp{\left ( {-v/20} \right )},\\
\beta_3&=&1\left /{\left\{ {\exp{\left ( {\frac{30-v}{10}} \right )}-1} \right \}}\right . ,
\end{eqnarray}
where $v$ represents the membrane potential, and $w_1{\mbox{ and }} w_2$ the activation and inactivation variables of the sodium current, and $w_3$ the activation variable of the potassium current.
The values of parameters are $v_{Na}=50\mbox{ [mV]},\ v_K=-77\mbox{ [mV]},\ v_L=-54.4\mbox{ [mV]},\ \overline{g}_{Na}=120\mbox{ [mS/cm$^2$]},\ \overline{g}_K=36\mbox{ [mS/cm$^2$]},\ \overline{g}_L=0.3\mbox{ [mS/cm$^2$]},\ {\mbox{ and }} C_m=1\mbox{ [$\mu$F/cm$^2$]}.$

\section{Definition of the matrices ${\bf A}_q$ and ${\bf B}$}{\label{matrixab}}

From {{{Eqs.}~{(\ref{rerespiketiming})}}},{{(\ref{defrf})}}-{{(\ref{i10deltaone})}}, and so on, ${\bf A}_q$ in {{{Eq.}~{(\ref{defab})}}} is written as
\begin{equation}
 {\bf A}_q=\left(\begin{array}{cccc}
-\frac{1}{c}{\frac{\partial {R}}{\partial{\left ( {{\delta\overline{t}_q}} \right )}}}&0&\ldots&0\\
-\frac{1}{c}{\frac{\partial {S_1}}{\partial{\left ( {{\delta\overline{t}_q}} \right )}}}&0&\ldots&0\\
\vdots&\vdots&&\vdots\\
-\frac{1}{c}{\frac{\partial {S_n}}{\partial{\left ( {{\delta\overline{t}_q}} \right )}}}&0&\ldots&0\\
\frac{-A_{{{\rm PP}}}\tilde{J}_{0q}{e}^{-{\left ( {\tilde{T}-q\tilde{T}/Q} \right )}/{\tau_{{{\rm PP}},1}}}}{cQ\tau_{{{\rm PP}},1}{\left ( {\tau_{{{\rm PP}},1}-\tau_{{{\rm PP}},2}} \right )}}&0&\ldots&0\\
\frac{A_{{{\rm PP}}}\tilde{J}_{0q}{e}^{-{\left ( {\tilde{T}-q\tilde{T}/Q} \right )}/{\tau_{{{\rm PP}},2}}}}{cQ\tau_{{{\rm PP}},2}{\left ( {\tau_{{{\rm PP}},1}-\tau_{{{\rm PP}},2}} \right )}}&0&\ldots&0\\
\frac{A_{{{\rm IP}}}{e}^{-{\left ( {\tilde{T}-q\tilde{T}/Q} \right )}/{\tau_{{{\rm IP}},1}}}}{cQ\tau_{{{\rm IP}},1}{\left ( {\tau_{{{\rm IP}},1}-\tau_{{{\rm IP}},2}} \right )}}&0&\ldots&0\\
\frac{-A_{{{\rm IP}}}{e}^{-{\left ( {\tilde{T}-q\tilde{T}/Q} \right )}/{\tau_{{{\rm IP}},2}}}}{cQ\tau_{{{\rm IP}},2}{\left ( {\tau_{{{\rm IP}},1}-\tau_{{{\rm IP}},2}} \right )}}&0&\ldots&0
\end{array}\right),\qquad q=1,\ldots,Q-1.
\end{equation}
In the same way, we obtain ${\bf B}$ in {{{Eq.}~{(\ref{defab})}}} as 
\begin{equation}
 {\bf B}=\left(\begin{array}{cccc}
-\frac{1}{c}{\frac{\partial {R}}{\partial{\left ( {{\delta\overline{t}_0}} \right )}}}+{\frac{\partial {R}}{\partial{\left ( {{\delta\overline{v}}} \right )}}}&{\frac{\partial {R}}{\partial{\left ( {{\delta\overline{w}_1}} \right )}}}&\ldots&{\frac{\partial {R}}{\partial{\left ( {{\delta\overline{w}_n}} \right )}}}\\
-\frac{1}{c}{\frac{\partial {S_1}}{\partial{\left ( {{\delta\overline{t}_0}} \right )}}}+{\frac{\partial {S_1}}{\partial{\left ( {{\delta\overline{v}}} \right )}}}&{\frac{\partial {S_1}}{\partial{\left ( {{\delta\overline{w}_1}} \right )}}}&\ldots&{\frac{\partial {S_1}}{\partial{\left ( {{\delta\overline{w}_n}} \right )}}}\\
\vdots&\vdots&&\vdots\\
-\frac{1}{c}\frac{\partial S_n}{\partial (\delta\overline{t}_0)}+\frac{\partial S_n}{\partial(\delta\overline{v})}&{\frac{\partial {S_n}}{\partial{\left ( {{\delta\overline{w}_1}} \right )}}}&\ldots&{\frac{\partial {S_n}}{\partial{\left ( {{\delta\overline{w}_n}} \right )}}}\\

\frac{-A_{{{\rm PP}}}\tilde{J}_{00}{e}^{-\tilde{T}/{\tau_{{{\rm PP}},1}}}}{cQ\tau_{{{\rm PP}},1}{\left ( {\tau_{{{\rm PP}},1}-\tau_{{{\rm PP}},2}} \right )}}&0&\ldots&0\\
\frac{A_{{{\rm PP}}}\tilde{J}_{00}{e}^{-\tilde{T}/{\tau_{{{\rm PP}},2}}}}{cQ\tau_{{{\rm PP}},2}{\left ( {\tau_{{{\rm PP}},1}-\tau_{{{\rm PP}},2}} \right )}}&0&\ldots&0\\
\frac{A_{{{\rm IP}}}{e}^{-\tilde{T}/{\tau_{{{\rm IP}},1}}}}{cQ\tau_{{{\rm IP}},1}{\left ( {\tau_{{{\rm IP}},1}-\tau_{{{\rm IP}},2}} \right )}}&0&\ldots&0\\
\frac{-A_{{{\rm IP}}}{e}^{-\tilde{T}/{\tau_{{{\rm IP}},2}}}}{cQ\tau_{{{\rm IP}},2}{\left ( {\tau_{{{\rm IP}},1}-\tau_{{{\rm IP}},2}} \right )}}&0&\ldots&0
\end{array}\qquad {\bf C}\qquad \right )
\end{equation}
with
\begin{equation}
{\bf C}=\left (\begin{array}{cccc}
{\frac{\partial {R}}{\partial{\left ( {{\delta I_{1}}} \right )}}}&{\frac{\partial {R}}{\partial{\left ( {{\delta I_{2}}} \right )}}}&{\frac{\partial {R}}{\partial{\left ( {{\delta I_{3}}} \right )}}}&{\frac{\partial {R}}{\partial{\left ( {{\delta I_{4}}} \right )}}}\\
{\frac{\partial {S_1}}{\partial{\left ( {{\delta I_{1}}} \right )}}}&{\frac{\partial {S_1}}{\partial{\left ( {{\delta I_{2}}} \right )}}}&{\frac{\partial {S_1}}{\partial{\left ( {{\delta I_{3}}} \right )}}}&{\frac{\partial {S_1}}{\partial{\left ( {{\delta I_{4}}} \right )}}}\\
\vdots&\vdots&\vdots&\vdots\\
{\frac{\partial {S_n}}{\partial{\left ( {{\delta I_{1}}} \right )}}}&{\frac{\partial {S_n}}{\partial{\left ( {{\delta I_{2}}} \right )}}}&{\frac{\partial {S_n}}{\partial{\left ( {{\delta I_{3}}} \right )}}}&{\frac{\partial {S_n}}{\partial{\left ( {{\delta I_{4}}} \right )}}}\\
{e}^{-\tilde{T}/\tau_{{{\rm PP}},1}}&0&0&0\\
0&{e}^{-\tilde{T}/\tau_{{{\rm PP}},2}}&0&0\\
0&0&{e}^{-\tilde{T}/\tau_{{{\rm IP}},1}}&0\\
0&0&0&{e}^{-\tilde{T}/\tau_{{{\rm IP}},2}}
\end{array}\right ).
\end{equation}

In our analysis, we have to evaluate the eigenvalues of ${\bf M}$ numerically, and hence the numerical evaluation of ${\bf A}_q$ and ${\bf B}$ is required.
The coefficients appearing in {{{Eqs.}~{(\ref{rfc})}}} and {{(\ref{sfc})}} are evaluated in Appendix~\ref{evaluationrs}.
Except for $b_{11},\ldots,b_{41}$, elements in ${\bf A}_q$ and ${\bf B}$ are determined by use of the values of coefficients obtained in Appendix~\ref{evaluationrs}, while we set $b_{11},\ldots,b_{41}$ so that ${\bf M}$ has the eigenvalue $\lambda_1=1$ with eigenvector~{{(\ref{solo})}}.

\section{Numerical evaluation of the coefficients in the functions $R{\left ( {{\left\{ {\delta \overline{t}_{q}} \right \}},\delta\overline{v},{\left\{ {\delta\overline{w}_{l'}} \right \}},{\left\{ {\delta I_{s'}} \right \}}} \right )}$ and $S_l{\left ( {{\left\{ {\delta \overline{t}_{q}} \right \}},\delta\overline{v},{\left\{ {\delta\overline{w}_{l'}} \right \}},{\left\{ {\delta I_{s'}} \right \}}} \right )}$ $(l=1,\ldots,n)$}{\label{evaluationrs}}

In order to evaluate the coefficients in the functions $R{[\ldots]}$ and $S_l{[\ldots]}\ (l=1,\ldots,n)$, we consider the single-body dynamics of the form
\begin{eqnarray}
\dot{v_0}&=&f{\left ( {v_0,w_{01},\ldots,w_{0n}} \right )}+\tilde{I}_0,{\label{vdynev}}\\
\dot{w}_{0l}&=&g_l{\left ( {v,w_{01},\ldots,w_{0n}} \right )},\qquad l=1,\ldots n
\end{eqnarray}
with
\begin{equation}
 \tilde{I}_0=\tilde{I}_{{{\rm PP}},0}+\tilde{I}_{{{\rm IP}}},
\end{equation}
where
\begin{eqnarray}
\tilde{I}_{{{\rm PP}},0}&=&\frac{A_{{{\rm PP}}}}{Q}\sum_{q'}\tilde{J}_{0q'}S_{{{\rm PP}}}{\left [ {t-t_{q'}^\ast(k)-\delta t_{q'}(k)} \right ]}+{\left ( {I_{1,0}^\ast{{\left [ {t_{{0}}^\ast({k})} \right ]}}+\delta I_{1,0}{{\left [ {t_{{0}}^\ast({k})} \right ]}}} \right )}{e}^{-{\left [ {t-t_q^\ast(k)} \right ]}/\tau_{{{\rm PP}},1}}\nonumber\\
&&+{\left ( {I_{2,0}^\ast{{\left [ {t_{{0}}^\ast({k})} \right ]}}+\delta I_{2,0}{{\left [ {t_{{0}}^\ast({k})} \right ]}}} \right )}{e}^{-{\left [ {t-t_q^\ast(k)} \right ]}/\tau_{{{\rm PP}},2}},\qquad t_0^\ast(k)<t<t_0^\ast(k+1)
\end{eqnarray}
and
\begin{eqnarray}
\tilde{I}_{{{\rm IP}}}&=&\frac{A_{{{\rm IP}}}}{Q}\sum_{q'}S_{{{\rm IP}}}{\left [ {t-t_{q'}^\ast(k)-\delta t_{q'}(k)} \right ]}+{\left ( {I_{3,0}^\ast{{\left [ {t_{{0}}^\ast({k})} \right ]}}+\delta I_{3,0}{{\left [ {t_{{0}}^\ast({k})} \right ]}}} \right )}{e}^{-{\left [ {t-t_q^\ast(k)} \right ]}/\tau_{{{\rm IP}},1}}\nonumber\\
&&+{\left ( {I_{4,0}^\ast{{\left [ {t_{{0}}^\ast({k})} \right ]}}+\delta I_{4,0}{{\left [ {t_{{0}}^\ast({k})} \right ]}}} \right )}{e}^{-{\left [ {t-t_q^\ast(k)} \right ]}/\tau_{{{\rm IP}},2}},\qquad t_0^\ast(k)<t<t_0^\ast(k+1).{\label{iipev}}
\end{eqnarray}
From {{{Eqs.}~{(\ref{i1})}}}-{{(\ref{i4})}}, we obtain the explicit form of ${\left\{ {I_{s,0}^\ast{{\left [ {t_{{0}}^\ast({k})} \right ]}}} \right \}}$ as follows
\begin{eqnarray}
I_{1,0}^\ast{{\left [ {t_{{0}}^\ast({k})} \right ]}}&=&\frac{A_{{{\rm PP}}}}{Q}\sum_{q'}\tilde{J}_{0q'}\frac{{e}^{-{\left ( {\tilde{T}-q'\tilde{T}/Q} \right )}/\tau_{{{\rm PP}},1}}}{{\left ( {\tau_{{{\rm PP}},1}-\tau_{{{\rm PP}},2}} \right )}{\left ( {1-{e}^{-\tilde{T}/\tau_{{{\rm PP}},1}}} \right )}},\\
I_{2,0}^\ast{{\left [ {t_{{0}}^\ast({k})} \right ]}}&=&\frac{A_{{{\rm PP}}}}{Q}\sum_{q'}\tilde{J}_{0q'}\frac{-{e}^{-{\left ( {\tilde{T}-q'\tilde{T}/Q} \right )}/\tau_{{{\rm PP}},2}}}{{\left ( {\tau_{{{\rm PP}},1}-\tau_{{{\rm PP}},2}} \right )}{\left ( {1-{e}^{-\tilde{T}/\tau_{{{\rm PP}},2}}} \right )}},\\
I_{3,0}^\ast{{\left [ {t_{{0}}^\ast({k})} \right ]}}&=&\frac{A_{{{\rm IP}}}}{Q}\sum_{q'}\frac{-{e}^{-{\left ( {\tilde{T}-q'\tilde{T}/Q} \right )}/\tau_{{{\rm IP}},1}}}{{\left ( {\tau_{{{\rm IP}},1}-\tau_{{{\rm IP}},2}} \right )}{\left ( {1-{e}^{-\tilde{T}/\tau_{{{\rm IP}},1}}} \right )}},\\
I_{4,0}^\ast{{\left [ {t_{{0}}^\ast({k})} \right ]}}&=&\frac{A_{{{\rm IP}}}}{Q}\sum_{q'}\frac{{e}^{-{\left ( {\tilde{T}-q'\tilde{T}/Q} \right )}/\tau_{{{\rm IP}},2}}}{{\left ( {\tau_{{{\rm IP}},1}-\tau_{{{\rm IP}},2}} \right )}{\left ( {1-{e}^{-\tilde{T}/\tau_{{{\rm IP}},2}}} \right )}}.
\end{eqnarray}
We solve the dynamics~{{(\ref{vdynev})}}-{{(\ref{iipev})}} under the condition
\begin{eqnarray}
v_0{{\left [ {t_{{0}}^\ast({k})} \right ]}}&=&v_0^\ast{{\left [ {t_{{0}}^\ast({k})} \right ]}}+\delta v_0{{\left [ {t_{{0}}^\ast({k})} \right ]}},\\
w_{0l}{{\left [ {t_{{0}}^\ast({k})} \right ]}}&=&w_{0l}^\ast{{\left [ {t_{{0}}^\ast({k})} \right ]}}+\delta w_{0l}{{\left [ {t_{{0}}^\ast({k})} \right ]}},\qquad l=1,\ldots,n.
\end{eqnarray}
Then, we obtain the functions
\begin{eqnarray}
v_0{{\left [ {t_{{0}}^\ast({k+1})} \right ]}}&=&F{\left ( {{\left\{ {\delta\overline{t}_{q'}(k)} \right \}},\delta\overline{v}_0{{\left [ {t_{{0}}^\ast({k})} \right ]}},{\left\{ {\delta\overline{w}_{0l'}{{\left [ {t_{{0}}^\ast({k})} \right ]}}} \right \}},{\left\{ {\delta I_{s',0}{{\left [ {t_{{0}}^\ast({k})} \right ]}}} \right \}}} \right )},\\
w_{0l}{{\left [ {t_{{0}}^\ast({k+1})} \right ]}}&=&G_l{\left ( {{\left\{ {\delta\overline{t}_{q'}(k)} \right \}},\delta\overline{v}_0{{\left [ {t_{{0}}^\ast({k})} \right ]}},{\left\{ {\delta\overline{w}_{0l'}{{\left [ {t_{{0}}^\ast({k})} \right ]}}} \right \}},{\left\{ {\delta I_{s',0}{{\left [ {t_{{0}}^\ast({k})} \right ]}}} \right \}}} \right )},\nonumber\\
&&\qquad l=1,\ldots,n.
\end{eqnarray}
It is straightforward to show
\begin{eqnarray}
&& R{\left ( {{\left\{ {\delta \overline{t}_{q}} \right \}},\delta\overline{v},{\left\{ {\delta\overline{w}_{l'}} \right \}},{\left\{ {\delta I_{s'}} \right \}}} \right )}\nonumber\\\
&=&\lim_{\epsilon\rightarrow 0}\frac{F{\left ( {{\left\{ {\epsilon\delta \overline{t}_{q}} \right \}},\epsilon\delta\overline{v},{\left\{ {\epsilon\delta\overline{w}_{l'}} \right \}},{\left\{ {\epsilon\delta I_{s'}} \right \}}} \right )}-v_0^\ast{{\left [ {t_{{0}}^\ast({k+1})} \right ]}}}{\epsilon},{\label{rfev}}\\
&& S_l{\left ( {{\left\{ {\delta \overline{t}_{q}} \right \}},\delta\overline{v},{\left\{ {\delta\overline{w}_{l'}} \right \}},{\left\{ {\delta I_{s'}} \right \}}} \right )}\nonumber\\
&=&\lim_{\epsilon\rightarrow 0}\frac{G_l{\left ( {{\left\{ {\epsilon\delta \overline{t}_{q}} \right \}},\epsilon\delta\overline{v},{\left\{ {\epsilon\delta\overline{w}_{l'}} \right \}},{\left\{ {\epsilon\delta I_{s'}} \right \}}} \right )}-w_{0l}^\ast{{\left [ {t_{{0}}^\ast({k+1})} \right ]}}}{\epsilon},\qquad l=1,\ldots,n,{\label{sfev}}
\end{eqnarray}
where $0<\epsilon\delta \overline{t}_{0}.$
The explicit value of $F{[\ldots]}$ and $G_l{[\ldots]}$ is easily computed by the numerical integration of the dynamics~{{(\ref{vdynev})}}-{{(\ref{iipev})}}.
We obtain $R{[\ldots]}$ and $S_l{[\ldots]}$ by evaluating $F{[\ldots]}$ and $G_l{[\ldots]}$ for sufficiently small $\epsilon.$

Once we obtain $R{[\ldots]}$ and $S_l{[\ldots]}$, we can easily evaluate the coefficients appearing in {{{Eqs.}~{(\ref{rfc})}}} and {{(\ref{sfc})}}.
For example, substituting ${\left ( {{\left\{ {0} \right \}},1,{\left\{ {0} \right \}},{\left\{ {0} \right \}}} \right )}$ into {{{Eq.}~{(\ref{rfc})}}}, we have
\begin{equation}
 R{\left ( {{\left\{ {0} \right \}},1,{\left\{ {0} \right \}},{\left\{ {0} \right \}}} \right )}=\frac{\partial{R}}{\partial{\left ( {\delta\overline{v}} \right )}}.
\end{equation}
Hence, $\frac{\partial{R}}{\partial{\left ( {\delta\overline{v}} \right )}}$ is calculated from $R{\left ( {{\left\{ {0} \right \}},1,{\left\{ {0} \right \}},{\left\{ {0} \right \}}} \right )},$ which is computed by {{{Eq.}~{(\ref{rfev})}}} with sufficiently small $\epsilon$.
In the same manner, we obtain every coefficient in {{{Eqs.}~{(\ref{rfc})}}} and {{(\ref{sfc})}}.

\label{lastpage}

{\newpage}

\end{document}